A Comparative Anatomy of REITs and Residential Real Estate Indexes:
Returns, Risks and Distributional Characteristics

by

John Cotter and Richard Roll

February 2011

Abstract

Real Estate Investment Trusts (REITs) are the only truly liquid assets related to real estate investments. We study the behavior of U.S. REITs over the past three decades and document their return characteristics. REITs have somewhat less market risk than equity; their betas against a broad market index average about .65. Decomposing their covariances into principal components reveals several strong factors. REIT characteristics differ to some extent from those of the S&P/Case-Shiller (SCS) residential real estate indexes. This is partly attributable to methods of index construction. Our examination of REITs suggests that investment in real estate is far more risky than what might be inferred from the widely-followed SCS series. REITs, unlike SCS series are forward looking, and this helps them in the prediction of SCS returns. REIT forecasts of SCS returns are reasonably precise over a number of periods.

|  | Authors' Coordinates | |
|---|---|---|
|  | Cotter | Roll |
| Address | University College Dublin Centre for Financial Markets School of Business Carysfort Avenue Blackrock, County Dublin Ireland | UCLA Anderson 110 Westwood Plaza Los Angeles, CA 90095-1481 USA |
| Voice | 353 1 716 8900 | 1 310 825 6118 |
| E-Mail | john.cotter@ucd.ie | rroll@Anderson.ucla.edu |

Acknowledgement

We have benefited greatly from discussions with Brad Case, Steve Cauley, Jerry Coakley, Joao Cocco, Craig A. Depken II, Stuart Gabriel, Joe Gyuorko, Cal Muckley, Brendan Murphy, Kathleen Smalley, participants at the 2011 American Real Estate and Urban Economics Society Annual Conference, and seminar participants at the University of Manchester and University College Dublin. We thank the Ziman Center for Real Estate at UCLA for providing REIT data and for a research grant supporting this project. Cotter acknowledges the support of Science Foundation Ireland under Grant Number 08/SRC/FM1389.

*A Comparative Anatomy of REITs and Residential Real Estate Indexes:*
*Returns, Risks and Distributional Characteristics*

I. Introduction.

Although real estate probably represents the dominant fraction of non-human capital for most Americans, it is very illiquid. Transaction costs and search costs are high when selling or buying a single- or a multi-family residence. Residential real estate indexes such as those published by S&P/Case-Shiller® (SCS) are widely followed, but they are reported only monthly and are subject to some unavoidable difficulties.[1]

The only truly liquid real estate related vehicle with high-frequency observability is the Real Estate Investment Trust or REIT. REITs are listed on major exchanges and are traded continually. Hence, their features should be of great interest to those who want to keep frequent track of their major investments. As proxies, REITs offer the opportunity to observe the likely behavior of all real estate prices if they were only observable. In this paper, we provide comprehensive documentation for REIT return characteristics and compare them to the Case-Shiller indexes of residential real estate.

Investors in real estate often are often motivated by a belief that unlike equities, real estate offers high returns and low risk possibilities. In a series of papers, Karl Case and Robert Shiller report tabulated responses from a questionnaire survey of home buyers (for example see Case and Shiller, 2003). Here large proportions of respondents suggest that their investment is based on large positive price movement possibilities that carry negligible risk. In order to address whether real estate is in fact a low risk and high return investment we empirically invest these characteristics of related price indexes using both SCS and REIT series.

We document a strong return performance for REITs in comparison to both the US stock market and, in particular, the SCS real estate series. Furthermore, we find that compared to thebroad-based equity index, the S&P500, REITs as a group has lower market risk (beta) but comparable total volatility. SCS risk features are negligible compared to the other series. REITs return distributions mostly display moderate negative skewness but very high kurtosis, which implies that they are emphatically non-normal, a result that we verify with formal tests. Their monthly returns are not very auto-correlated. These characteristics are similar to those noted for equity returns.

---

[1] Later in the paper, we document some of the measurement difficulties for these indexes and analyze their consequences.

In contrast, SCS index returns are highly auto-correlated, probably because they are constructed as three-month moving averages. They also have much smaller total volatility, roughly one-fifth that of REITs. Reconstructed REIT series as a three-month moving average leads to a far greater alignment of the risk/return outcomes between the two sets of real estate series. This reconstruction reduces volatility dramatically and increases autocorrelation , thus suggesting that the unadjusted REITs are much more in line with the performance of actual real estate markets. However, SCS returns like REITs, are moderately left-skewed and non-normal, but their betas are very small, only 1/20 as large as the betas of REITs (again resulting to an extent from the construction method followed by the SCS series).[2] We report further divergences in the real estate indicators, where for instance, to explain 90% of REIT return variance, five to six principal components are required. In contrast, only three principal components explain 90% of the variance in SCS index.

Much of the recent financial crises had focused on the pivotal role of real estate and its poor performance. Various explanations including the sub-prime mortgage meltdown, inadequate financial legislation, increased availability of mortgage credit, and a real estate bubble heavily inter twine with the collapse in real estate markets (for a discussion of these and other drivers of the crises see Mian and Sufi (2009) and Roll (2011)). Real estate is a fundamentally important asset class and represents a large proportion of the net worth of individuals, firms and the overall economy.[3] Whilst detailing past performance can help us understand the role of risk in the crises, the ability to predict future real estate can help us avoid repeating past failures. It is thus important to determine if we can predict future real estate returns. Taking our two surrogates of real estate we examine their predictability on a stand-alone basis and in conjunction with each other. The forecast from the smoothed SCS series will shrink large possible price movements of future real estate prices in comparison to REITs. But is this how real estate prices actually behave? Due to the lack of observability of true prices, this is a question that cannot be directly answered. However, we can determine if the indicators can help in predicting future values of each other, and more importantly, identify which series has a better ability in doing so. Thus we can see if REITS can help in predicting the often cited and heavily used SCS series. We can also measure the extent that both series respond to innovations in each other and whether this can aid forecasting future real estate returns.

---

[2] Although Scholes-Williams (1977) betas are somewhat larger, (e.g., for Los Angeles the beta increases from 0.027 to 0.0624), they remain small in comparison to those of REITS.
[3] For instance, in 2007, the value of U.S. residential real estate totalled US$ 22.5 trillion compared to US$ 19.9 trillion in domestic equities.

In addressing these issues, we find evidence of real estate predictability using the REIT series. The forward looking investor orientated REIT series can predict future SCS index returns. We find much of the future variation of SCS returns can be predicted by the inclusion of movements in the REIT series. Thus, adding REIT series help in forecasting future real estate returns as proxied by the SCS series. We also report evidence of Granger causality between the series on a stand-alone basis suggesting that, for example, changes in REITs in the current period impacts changes in the SCS series in future periods. Moreover, REITs help to offer reasonable precise forecasts of SCS series over a number of periods. As the SCS series is developed using a moving average construction we report that predictability of itself is strong and remains for long lags compared to the REIT series. Notwithstanding this, both impulse response and forecast error variance decomposition statistics suggest that innovations in REIT series have additional information for forecasting future SCS returns.

These and other details are reported numerically below after we first describe the data sources and provide summary statistics covering the two separate indicators of real estate investments. We begin however with a discussion of the literature on our proposed proxy for real estate, REITS, detailing their risk characteristics and evidence on predictability.

II. Literature Review

There is considerable literature focusing on the risk profiles of REITS. Chan, Hendershott and Sanders (1990) use the macro-economic factors identified for equities by Chen, Roll and Ross (1986) and find that changes in risk, term structure and unexpected inflation drive both REIT and equity returns. Peterson and Hsieh (1997) explain REIT returns with the three Fama-French (1993) factors and Derwall, Huij, Brounen and Marquering (2009) add a momentum factor. See also Lizieri, Satchall and Zhang, (2007). However, there is some indication that broad equity factors are becoming less important for REITS and are being replaced by specific real estate factors (Clayton and MacKinnon, 2003).

Idiosyncratic risk is important for real estate in general and for REITs in particular. Investors in real estate tend to hold small undiversified portfolios due to the localized and segmented nature of the asset. Clayton and MacKinnon (2003) find that idiosyncratic risk for REITS is large and increasing over time. Ooi, Wang and Webb (2009) find that there is a positive risk and return relationship associated with idiosyncratic risk of REITs so the expected returns of real estate investors increase with idiosyncratic risk. Moreover REITS tend to have low levels of systematic risk, although this varies across assets (Gyuorko and Nelling, 1996; Peterson and Hsieh, 1997).

REITs indexes are forward looking as they represent investor behavior relating to real estate markets. Real estate returns by themselves, especially commercial, are mean reverting and follow cycles, and are found to be predictable. REITs have strong predictive power for future REITS (Chui, Titman and Wei, 2003). However, there is mixed to weak evidence that REITs prices can predict future real estate prices. This is somewhat surprising as REITs offer liquidity and transparency for real estate investors who through their trading actions should process and signal information about future values of real estate. In a similar vein, the forecastabilty of equity returns is mixed (for a review see Goyal and Welch, 2007).

Of the evidence in support of REIT predictability, Gyuorko and Keim (1992) find that lagged values of REIT returns can predict returns on appraisal-based real estate indexes. They note that appraisal-based indexes incorporate market fundamentals at a lag. Analogously the use of past moving average based index values drive the development of transaction based series such as Case Shiller. However REITs were unable to predict the decline in real estate during the recent crises, and Pavlov and Wachter (2009) suggest that this may be due to cheap credit being offered and accessed by REIT investors.

Appraisal based methods (similar to SCS series) are heavily reliant on smoothing based techniques and there is a substantial divergence between these returns and REITS (Muhlhofer, 2008). Since REITs are not subject to the dampening of volatility associated with real estate appraisals (Ross and Zisler, 1991) of the moving average based SCS series, an excellent case can be made that REITs are among the best available short-term proxies for real estate generally, even residential real estate.

REIT returns are non-normal and exhibit volatility clustering. Young and Graff (1995) find support for a mixtures distribution where the variance is not constant in contrast to the normal distribution. Lizieri, Satchall and Zhang (2007) identify and model excess kurtosis in REIT returns. (Excess kurtosis is related to fat-tails where the probability distribution for large returns exceeds that of the normal distribution.) Booth and Broussard (2002) find support for fitting the fat-tailed Fréchet distribution to REIT returns. In line with equities REITS returns exhibit conditional heteroskedasticity and volatility has been successfully modeled with many GARCH models (see long memory modeling with FIGARCH by Cotter and Stevenson, 2008).

III. Data

Monthly return data and information about investment strategies are available for individual REITs from the UCLA Ziman Center for real estate. This information is also available now on the CRSP database from the Center for Research in Securities Prices at the University of Chicago. The data are available daily since 1969 but since we want to compare REITs and the SCS indexes, our analysis will be based mainly on monthly observations. As we are using relatively low frequency REIT data we avoid problems of thin trading (although the CRSP/Ziman Centre database incorporates used prices to overcome this for daily data).

REITs come in various flavors. As reported in Table 1, there are three broad categories, equity, mortgage, and hybrid. Equity REITs are devoted to direct purchases of real estate requiring at least 75% of their total assets in income producing real estate properties. Mortgage REITs hold portfolios of loans backed by real estate collateral with at least 75% of their assets in residential mortgages, short and long term construction loans and mortgages in commercial properties. Hybrid REITs are a combination of the two, investing in both properties and real-estate related loans. Within these categories, the REITs are further categorized by the main property types held or financed such as residential property.

Our sample includes all REITs that have traded on the NYSE, AMEX and NASDAQ exchanges during this time. Table 1 also reveals the enormous growth of the REIT sector; from the beginning of 1980 through May, 2009 the total market capitalization of REITs had grown 80-fold, almost 17% per year (compounded) over more than 28 years. This growth has been in assets and in the number and aggregate market value of equity REITs. The numbers of mortgage and hybrid REITs have actually declined as well as their relative market value. Equity REITs represent over 90% of total value at the end of the sample and hybrid REITs now have negligible size.

REITs are generally highly leveraged assets although there is considerable variation across REIT types. Overall they tend to have higher debt than ordinary firms. Mortgage REITs have higher debt than equity REITS. Leverage levels tend to be strongly related to the ratings of the debt with an increase in the number of investment grade debt offerings resulting in higher levels of debt. Jaffe (1991) offers further explanation for the high levels of REIT debt. He suggests REITs tend to be relatively small firms but their cost of debt is lower than similarly-sized ordinary firms. The investments of REITs in real property, which is associated with a high level of debt capacity, leads to more debt financing.

Over time, a number of legislative acts have responded to the changing investment environment and have allowed REITS more flexibility to meet new challenges.[4] Although REITS were introduced initially as an investment vehicle that allowed and encouraged small investor involvement, the role of institutional investors, and especially pension funds, has increased over time. For instance, prior to 1980, 50% of REIT shareholding could not be held by groups of five or fewer individual investors, known as the 5/50 rule. However, these rules were relaxed over time and the average institutional ownership increased from 10% to 39% between 1981 and 2009. This included the Omnibus Budget Reconciliation Act (1993) that allowed pension funds to overcome the 5/50 rule (prior to the legislation they were treated as a single investor) by counting their own investors as individual investors. As a consequence REITs capitalization grew in the 1990s accompanied by increased interest from institutional investors.

REITS as a real estate investment vehicle offer some advantages. In contrast to other entities, they are effectively exempt from corporate taxes. However, there are other rules that may offset this tax benefit. REITS must distribute a high proportion, usually over 90%, of their taxable income. This restricts their ability to grow from internal funds. Also, some REITS that have been set up with pre-determined finite lives; these generally rely less on additional external funding and tend to have limited growth.

REITS developed a pronounced tendency to use external advisors during the 1980s to manage their assets. This is probably inherited from the Real Estate Investment Trust Act (1960), the original legislation that introduced REITs, which defined REITs as having multiple trustees as managers.

Perhaps the most confining restriction is a limit on the type of income a REIT can earn and the type of asset it can hold. But despite such limits, REITs have been generally profitable since the inception.

III.A. Descriptive Statistics.

---

[4] These include the REIT Modernisation Act (1999) and the Economic Recovery Act (1981); (see Chan, Erickson and Wang [2003] for a review).

Table 2 reports summary statistics for individual REITs by calendar year from 1980 through 2008. Although mean returns are positive on average, there are substantial cross-sectional differences among individual REIT returns in every calendar year. The cross-sectional standard deviation of individual REIT means – the mean of the standard deviation of returns by calendar year - is generally six to ten times larger than the mean itself. On average, betas are less than unity, thereby indicating smaller systematic risk than equities in general. Again, however, there is substantial cross-sectional variation. Also, it should be noted that betas <u>exceed</u> 1.0 in each of the last five sample years. This might be due to equity REITs becoming larger in number relative to presumably less risky mortgage and hybrid REITs. The betas for Equity REITs are greater than one for the last five sample years and are larger than corresponding values for Hybrid and Mortgage REITs (with the exception of 2004 where beta for Mortgage REITs is larger).

To study the general characteristics of REITs and REITs in various categories, we form each month several value-weighted portfolio of individual REIT returns weighted by the market capitalization at the end of the prior month. One index covers all REITs, three others are for the broad equity, mortgage and hybrid groups, and six others were composed of specific property types. The first observation is for January, 1980 and the last available observation is in May, 2009.[5]

The S&P/Case-Shiller® (SCS) Indexes were collected from their inception in February 1987 through May 2009. SCS indexes are created by combining all transactions on a quality-adjusted basis for repeat sales on single family homes (for full details see Standard and Poors, 2009). Many forms of real estate are explicitly excluded, for example condominiums and multi-family dwellings and this may affect the risk and return performance of the indexes. What will certainly dampen return and volatility statistics is the downward adjustment that is followed for extreme price movements, compared to average market price movements.[6] The price change included in the index is for two arms-length sales of the same home. The SCS indexes are reported monthly with each index point representing a three-month moving average of the current and past two months values. Given the lack of observable real estate prices at high frequencies the three-month moving average is an attempt to have a sufficiently large sample size with meaningful price changes.[7]

---

[5] The returns reported are nominal but excess returns and inflation-adjusted returns provide a similar picture. They are available from the authors upon request.

A value-weighted index that operates in an analogous fashion to a market capitalization weighted equity index is created that distinguishes the real-estate unit traded across three price tiers, low, medium and high. Case and Shiller (1994) have noted that there is a value effect across the three tiers with low tiers having better return performance compared to the other tiers. They compare this to similar size based investment strategies followed in equity markets. There were 14 city indexes available over the entire time period, 1987-2009[8] and a "National" index constructed as an aggregation of 10 of the major metropolitan areas. In addition to the 10 cities (Boston, Chicago, Denver, Las Vegas, Los Angeles, Miami, New York, San Diego, San Francisco, Washington DC) representing a ten-city composite, data is also available for Charlotte, Cleveland, Portland and Tampa.

Table 3 presents descriptive statistics for both our set of REIT indexes and for the SCS indexes.[9] Time series plots of a sample of the series returns and conditional volatility (using a GARCH (1, 1) model) are given in Figures 1 and 2. Except for the Lodging Resort category, REIT returns are, on average, somewhat higher than the returns of the SCS indexes. This is particularly important because means are virtually immune to such problems as moving average-induced autocorrelation and spuriously low volatility. That is, over a long time period, the general drift in prices should be a valid indication of investment performance, regardless of the nature of short-term fluctuations. For example, the residential REIT index has a mean monthly return of 1.005%, slightly more than 12% per annum, while the SCS composite has a monthly mean of 0.372%, slightly more than one-third as large. This is something of a puzzle. In addition to the diverging magnitudes of returns for the respective real estate series, the smoothness of returns for the SCS series resulting from the moving average construction is further evidenced in Figure 1.

---

[6] Standard and Poors (2009) provide illustrations of cases where the weighting of extreme price changes would be adjusted downwards, including transactions for homes that were not well maintained and thus deviating from a representative market real estate unit.

[7] There are many other adjustments in the creation of the indexes to try and ensure meaningful prices. For example, if a transaction occurs too quickly on a unit they can be removed if they are not felt to represent market prices. The cut-off point is six-months. In contrast, if the arm-length transaction was over a very long holding period they would be a given a lower weight in the database than a transaction over a shorter (six-month) period (the interval adjustment process is further detailed in Case and Shiller, 1987).

[8] Twenty SCS cities are now included in their indexes, but only since 2000.

[9] REIT results are presented first by REIT type and then by Property type. For REITS a condensed set of property types outlined in Table 1 are analysed as there is no price data available for Unknown property type and Mortgage property type. Moreover data is only available for Health Care property type since March 1984 and for Self Storage property type since November 1982.

REIT volatilities are much larger than SCS volatilities. Figure 2 illustrates the divergence in volatilities especially during the recent bear market. The ratio of volatility for All REITs relative to the SCS composite is 4.916/.907 = 5.42 for the full sample. This is a large difference but is partly attributable to the moving average construction method of SCS. Clearly, the dramatically larger auto-correlation of the SCS returns have arisen, at least in part, for the same reason. To illustrate, applying the SCS method of a 3-month moving averages to the All REIT series brings a reduction in volatility to 2.871. Moreover, the auto-correlation of the moving average of the All REIT series increases to 0.712, similar to the SCS series.[10]

The All REITs return has a beta of 0.625 against the S&P500, thereby indicating considerably smaller systematic market risk. Most REIT sub-categories are similar; indeed, their betas are all in the 60% range with the exception of Hybrid REITs (0.511) and Lodging Resort REITs (0.824). In sharp contrast, SCS betas are tiny, but again, this is attributable to some extent by their construction. For instance the beta from using a moving average of the All REIT series reduces to 0.211.

Except for Lodging Resort, all the REIT indexes have a moderate amount of negative skewness. This characteristic is shared by every one of the SCS indexes, one of the few statistics between the two sets of real estate indicators that are more or less in agreement. Kurtosis is moderately higher for REITs than for the SCS indexes, which is also reflected in the Jarque-Bera tests of normality. Every series is significantly non-normal and all have the thick tails typical of most asset returns.

---

[10] Other possible causes of the divergence between the outcomes of REIT and SCS series are the different forms of leverage associated with investing in REITs compared to investing in single-family properties that constitute the SCS series. REITs would tend to have higher leverage amplifying the magnitude of price movements during boom and crises periods.

III.B. Principal Components

Using the six REIT index series and separately the fourteen SCS series, we extract principal components from the covariance matrix of their concurrent monthly returns.[11] Figures 3 and 4 present "scree" plots for the narrow (disaggregated) set of REITs and SCS indexes, respectively. A scree plots depicts the variance explained by successive principal components (PCs) ranked from left to right in descending order of variance explained. The bar gives the variance explained by the PC and the cumulative variance explained in percent is printed above each bar.

Comparing the figures, these two sets of real estate indicators again display some differences. The first PC explains quite a bit more of REIT returns than it does for SCS returns; the same is true for the second PC. Moreover, it takes six PCs to cumulatively explain about 90% of SCS volatility while only three PCs are required to reach the same level of cumulative explanation for REITs. One possible explanation is that REITs are less heterogenous than residential real estate measured by the SCS indexes. The SCS indexes are geographically more diverse than REITs, which mix in properties from all over the country. On the other hand, REITs do include commercial properties of various types and also mortgages. So the empirical contrast is far from a foregone conclusion.

Table 4 gives the factor loadings on the first 2 PCs for the separate indexes and the variance explained by the first 3 PCs for REITS and first 6 PCs for SCS indexes. The number of factors is chosen to explain 90% of the variance of the series. Across the REIT indexes, there is only minor variation in the first factor loading and in explaining the variance. The SCS numbers show more variation across cities. For example, Charlotte has only a 0.090 loading on the first PC while San Francisco's loading is 0.388. Charlotte's R-square of 0.566 (and its relatively small loading on the first PC) seems to suggest that its real estate market is partially segmented from the markets in other urban centers.

---

[11] Principal Component Analysis (PCA) is actually completed twice. In the first case, we use all series to obtain residuals after removing common factors, thereby allowing for an analysis of the idiosyncratic risk associated with each series. Secondly, we examine the structure of the returns with PCA applied to a narrower set of series that excludes any aggregated series such as the SCS Composite 10. Except for the discussion involving residuals, we are referring to the latter PCA.

Correlations between these REIT factors and various candidate variables raises hope that the factors can be identified. There is a strong and significant correlation between the first REIT factor and the general stock market (corr = 0.591) as proxied by the S&P500 index. In contrast, the first SCS factor series is more related to industrial production growth (corr = 0.311), than to the stock market (corr = 0.170). The factors ($2^{nd}$ for REITs and $4^{th}$ for SCS) for both sets of series respond negatively to changes in interest rates proxied by the 3-month US Treasury bill rate. The impact of interest rates is stronger for the SCS series (corr = -0.356 compared to corr = -0.194 for REITs).

III.C. Risk of Loss Measures.

We now turn our attention to risk of loss measures. Value-at-Risk is a widely used indicator of loss likelihood. It is mandated in some countries for financial reporting by institutions such as banks, in which case it covers the entire asset/liability portfolio. It can also be adapted to individual asset series such as our REIT indexes and the SCS residential indexes.

To obtain a risk of loss measure, one must first fit a probability distribution to the empirical series. When Value-at-Risk was first used, the Gaussian or normal was commonly employed as the fitted distribution; it is convenient since it is fully characterized by just two parameters, the mean and variance, which can be estimated easily from a sample of observations.

We now know, however, that the normal distribution is a poor model for most financial asset returns and is particularly prone to large errors in Value-at-Risk. The main reason is that asset returns often have thicker tails than the normal and hence larger probability of extreme loss. As we have already seen, REITs and SCS depart significantly from the normal in this respect. Consequently, we resort here to three alternative models of the return distribution based on fitting the actual historical observations.

The first model will be called EM for the Efficient Maximization algorithm fit to the unconditional distribution. As shown in Table 3, excess kurtosis is present in all of the series. This characteristic is typical of Gaussian mixtures with extensive regimes of small returns interspersed with occasional extremely large returns, thus giving rise to fat-tails. Each Gaussian distribution in the mixture would have its own mean and standard deviation while a fifth parameter is probability of being in one regime or the other (see Hamilton, 1994, pp. 685-689). The EM algorithm uses maximum likelihood estimates to determine the parameters (Dempster, Laird and Rubin, 1977) and the Bayesian Information Criterion (BIC) to determine model selection. In all cases except 4 (San Diego, Washington, Miami and Tampa) two mixture were identified as the optimal model incorporating regimes of low (high) volatility with a high (low) probability of occurrence.

The second model is the Generalized Pareto Distribution (GPD), which is often used to fit thick-tailed empirical phenomena. The GPD relies on Extreme Value Theory. We adopt a Peaks over Threshold (POT) approach for its fit; (See Embrechts, Kluppelberg and Mikosch, 1997).[12] This approach utilizes the realizations of a random variable *X* in excess of a high threshold *u*; such a realization is called an "exceedence." When *u* is large, as it would be for tail realizations of financial time series, the distribution of exceedances tends to a GPD. The GPD parameters, the shape and scale parameters of exceedences, are estimated using maximum likelihood. The GPD has proven successful in modeling the fat-tails characteristics for a number of financial instruments such as currencies and equity returns (e.g., Cotter and Dowd, 2006).

The third model is based on the GARCH (1,1) (Bollerslev, 1986) fit to non-stationary volatility in the return series. It makes some sense to fit a GARCH model to our series because non-stationarity in the parameters, particularly the variance, is often given as an explanation for the appearance of thick tails. For example, it is easy to show that a mixture of Gaussian distributions that cycles through several volatility regimes will have an unconditional thick-tailed distribution. Also we find significant ARCH effects for all series reported in Table 1.

---

[12] Alternatively, extreme tail returns could be modelled by Generalised Extreme Value (GEV) theory, which deals with the distribution of the sample maxima. The GEV and POT approaches are analogous in the limit, but we prefer to use the POT approach because it (generally) uses one less parameter, and because the GEV approach does not utilise all extreme returns if extremes occur in clusters.

GARCH differs from EM in that it models some persistence in the conditional volatility. EM, in contrast, assumes that the volatility regime shifts randomly from period to period.

For each of these three models, we fit the empirical sample and derive distributional parameters (which are not reported for reasons of space.) Then, the risk of loss is calculated by examining the left tails of the fitted distributions and parameters.

We report two loss statistics. The first is simply the fractile of the loss distribution; i.e., it is the loss that is met or exceeded with a particular probability such as 1%. The 1% probability of loss is the 99% fractile of the fitted loss distribution in the sense that 99% of the time the return will be larger.[13] The second statistic is the average loss given that a particular loss fractile has been breached. This is the expected value in the left tail of the return distribution conditional on the return being lower than a particular fractile.

Table 5 has the results. The numbers there are percentage <u>losses</u> so they are the negatives of the monthly realized returns. For example, the Generalized Pareto Distribution (GPD) produces an estimate of -27.539 percent for the average loss in the All REITs index given that its return has been less than or equal to -21.658%, the .999 loss fractile.

Thus, according to GPD, the All REITs index return will be less than -21.658% one month in every thousand (83 years and 4 months) and when this happens the average return will be -27.539%  Table 3 shows that the mean monthly return of this index is 0.884% and its return standard deviation is 4.916%. Hence, a loss of -21.658% is 4.4 standard deviations below the mean. It should be noted that the .001 fractile of the Normal distribution is only 3.09 standard deviations below the mean, which would imply a return of "only" -14.3%. The reality is much worse than that depicted by an assumption that returns are Normally distributed. Note also the EM results tend to be small relative to the GPD and GARCH estimates. This is a result of the parameters of the mixtures distributions where the high volatility distribution has a very low probability of occurrence and results in the overall risk measures having lower magnitudes.

---

[13] The fractile of the loss distribution is 1 minus the fractile of the return distribution.

Pretty much the same story can be told for all the REIT indexes and for the SCS indexes as well (though the latter, of course, have much lower volatility.)

Much of the volatility in real estate is explained by broad market factors; indeed, as we have already seen in Table 4, principal component factors explain 80 to 90 percent for both REIT and SCS indexes, with a few exceptions among the latter. However, individual real estate assets would also be exposed to idiosyncratic risk, particularly for many families whose home is the main, and undiversified, investment. Consequently, it is worthwhile to examine the distributions of residual or idiosyncratic risks, after common variation is excluded. To this end, we first calculate the residuals after regressing REIT returns on three PC factors and SCS returns on six PC factors, (based on the findings above.)

The remaining residual risk is important for many reasons. In the future, it may be possible to eliminate common variation using real estate futures. Even now, long short strategies commonly employed by hedge funds can be engineered to eliminate systematic risk. But residual risk remains in both cases.

The issue we examine here is the impact of the shape of tail distributions on real estate residual risk. This would be relevant, for instance, in computing value-at-risk for a hedge fund that follows a tailored long/short strategy. The same methods can be employed. Table 6 has the results.

To illustrate with an example, let's take the Unclassified REIT index, which Table 6 reports has a Generalized Pareto Loss Distribution .95 fractile of 3.462. Table 3 shows that this index has a total return standard deviation of 5.242 and a mean of 0.730 while Table 4 reports that its R-square on the PC factors is .903. The residuals' standard deviation is thus $5.242(1-.903)^{1/2}$ or approximately 1.63, so the .95 fractile in Table 6 is $3.462/1.63 = 2.12$ standard deviations below the mean residual (which is zero.) Five percent of the months, Unclassified REITs will have idiosyncratic (non-market) returns less than 3.462 percent but the Normal distribution would have indicated losses of only $1.65(1.63) = 2.69$ percent. Again, as with raw returns, residual returns have thick tails relative to the Normal distribution although the magnitude of the risk is considerably smaller.

III.D. Return Predictability

Given that REIT prices represent investor trading activity in real estate assets, it is interesting to ascertain whether they have any predictive power for future real estate prices. The liquidity and availability of high frequency forward looking investor prices may have predictive power. As we have already seen, REITs and SCS are correlated, and especially if one constructs the REIT series in the same manner as followed by the SCS series.

To motivate the predictability analysis, we show plots of the time series of a selection of real estate log price series in Figure 5 between 1987 and 2009. We use an MA(3) of the REIT series to align it with adjustments made in the development made in constructing the SCS series although the conclusions are equally relevant where no adjustment is made. The positive performance of all real estate series is evidenced where a similar pattern occurs for SCS and REIT series. Much of the sample has an upward trend for all series that appears to begin earlier for REITs, followed by a sharp decline in real estate prices in recent times.

To avoid a spurious regression from these similar trends we fit a VAR to the returns data and results are reported in Table 7. In Panel A we confirm that the log price series exhibit a unit root whereas log returns are stationary. Next we report that REITs have predictive power for other real estate series. Taking the Residential REIT series as an example, we see from the VAR model coefficients in Panel B that it has predictive power for the SCS series at lag 3 as well as having forecasting power for itself. Similarily the SCS series is able to predict the REIT series. The VAR fit indicates strong explanatory power with over 90% of the variation in SCS returns being forecastable ahead of time with the REIT series. Also we seea good fit for the VAR between actual and response values in Figure 6, and especially for the SCS series. We report model coefficients from a VAR (5) chosen using both the AIC and BIC criterion although we also examined different lag specifications of the VAR and similar results are observed. It is clear that the All REITs series also has predictive power for the SCS series.

We also examine causality. To motivate the causality, tests we examine autocorrelations and cross correlations of the REIT and SCS series. An illustration for the Composite10 and two REITs series, the Residential and All REIT series are given in Figure 7. Whilst all series are positively autocorrelated, significant effects remain at longer lags, more than 20, for the SCS series. This is to be expected given the moving average construction of the SCS series. Both REITs show, however, lesser predictability in themselves but follow a similar pattern to each other. The cross autocorrelations provide evidence of correlation between the Composite10 and lags of the REIT series. Both residential and All REIT series are significantly impacted by movements in the SCS series and vice versawith a positive sign for a number of lags,.

Formally testing for the REIT series being able to help in forecasting the SCS series we find evidence of Granger causality. In Table 7, panel C, we report significant F-statistics that suggest that we should not reject the null hypothesis that the SCS series with its lags have an ability to forecast the REIT series and similarity for the null that the REIT series forecasts SCS returns at conventional levels. This direction is bi-directional suggesting feedback between the proxy real estate series.

In panel D of Table 7 we present multi-period forecasts of the respective returns series. Lags of REITs and the SCS series are able to aid the forecasting of the other real estate surrogate series for a number of periods. The forecast returns for all real estate indicators are less that the mean returns reported in Table 3 for most periods. For example, the return of 0.230% at the beginning of the forecast horizon is less than 0.327% for the SCS series in Table 3. Overall, the VAR gives us reasonably precise forecasts of future SCS series although the standard errors of the REIT series forecasts are relatively much larger.

We also look at the evidence of impulse response functions and forecasted variance decomposition for forecasting real estate returns. The impulse responses suggest that both REIT and SCS series respond positively to innovations in the other series. Impulse response functions are presented in Figure 8 for the VAR where we order the variables as Composite10 followed by the REIT series (although ordering has no qualitative effect on responses). For both the REIT and Composite 10 series, there are significant responses to innovations in the other series and these remain for approximately six months. The pattern in impulse responses is similar for both REIT series. We also see that the own-series response to innovations for SCS is stronger than both REIT series with both a larger effect and one that decays more slowly, significant for twenty four months.

To see how much of the forecast of the future error variance is explained by innovations in the real estate series we utilize our VAR output and present these in Figure 9. Whilst, the forecast error variance decomposition suggests that the main response of both REIT and SCS series is from the orthogonal innovations in the respective series, there is also a considerable fraction explained by innovations in the other series. Innovations in both sets of series have explanatory power of over 5% for the multi-step forecast error variance of the other series within a few lags with a similar pattern of results for both REIT series. Thus, overall we report a large spectrum of results that support predictability in real estate indicators.

Conclusion

We study return data for U.S. Real Estate Investment Trusts (REITs) over the past three decades. REITs have somewhat less market risk than other equity classes; their betas against a broad market index average about .65. In contrast, the more commonly used real estate indicator, the S&P/Case-Shiller (SCS) residential real estate indexes have a much lower beta, although this in part is due to its moving average construction.

REIT characteristics differ to some extent from those of the SCSindexes. Broad REIT indexes are about five times more volatile than the SCS indexes and have three times higher returns on average. The associated risk of loss measures for REITS are also considerably higher than for the SCS series. Also, unlike SCS returns, REIT returns exhibit have very little autocorrelation. Extracting principal components from REIT and SCS returns reveals another difference; six factors are required to explain 90% of the volatility in SCS returns while only three factors are required for REIT returns. These distinguishing features must be partly attributable to differing methods of index construction.

REITs prices are forward looking and constructed based on investment transactions whereas in contrast, SCS series are obtained using a moving average of previous index values. In turn, the REIT series returns have predictive power for the SCS series. Inclusion of REITs improves the forecastability of the commonly cited SCS series. Forward looking REIT returns provide reasonable precise forecasts of future SCS returns over a number of periods. Moreover, there is causality between the series, suggesting that analysis of current REIT return values is beneficial for predicting future SCS returns.

# References

Bollerslev, Tim, 1986. Generalized autoregressive conditional heteroskedasticity, Journal of Econometrics, vol. 31(3), pages 307-327.

Case, Karl E. and Robert J. Shiller 1987, Prices of Single-Family Homes Since 1970: New Indexes for Four Cities, New England Economic Review, September/October, 46–56.

Case, Karl E. and Robert J. Shiller, 1994, A Decade of Boom and Bust in Prices of Single Family Homes: Boston and Los Angeles, 1983 to 1993, New England Economic Review, March/April, 39-51.

Case, Karl E. & Robert J. Shiller, 2003. Is There a Bubble in the Housing Market?, Brookings Papers on Economic Activity, Economic Studies Program, 34(2003-2), pp 299-362.

Chan, Su Han, John Erickson, and Ko Wang, 2003, Real Estate Investment Trusts: Structure, Performance and Investment Opportunities, New York: Oxford University Press.

Chan, K. C., Patric H. Hendershott and Anthony B. Sanders, 1990, Risk and Return on Real Estate: Evidence from Equity REITs, AREUEA Journal, Vol. 18, No. 4, 1990

Chen, Nai-Fu, Richard. Roll and Stephen. A. Ross. 1986, Economic Forces and the Stock Market: Testing the APT and Alternative Asset Pricing Theories. Journal of Business 59, 383-403.

Chui, Andy C.W., Sheridan Titman, and K.C. John Wei, 2003, The Cross-Section of Expected REIT Returns, Real Estate Economics, 31, 451–479.

Clayton Jim and MacKinnon Greg, 2003, The relative importance of stock, bond and real estate factors in explaining REIT returns, Journal of Real Estate Finance and Economics 27, 39-60.

Cotter, John, and Kevin Dowd, 2006, Extreme Spectral Risk Measures: An Application to Futures Clearinghouse Margin Requirements, Journal of Banking and Finance, 30, 3469-3485.

Cotter, John, and Simon Stevenson, 2008, Modelling Long Memory in REITs, Real Estate Economics, 36, 533-554.

Dempster, A. P., N. M. Laird, D. B. Rubin, 1977, Maximum likelihood from incomplete data via the EM algorithm, Journal of the Royal Statistical Society, Series B, vol. 39, pages 1-38.

Derwall, Jeroen., Joop Huij, Dirk Brounen, and Wessel Marquering, 2009, REIT Momentum and the Performance of Real Estate Mutual Funds, Financial Analysts Journal, 65, 5, 24-34

Embrechts, Paul., Kluppelberg Claudio., Mikosch, Thomas., 1997, Modelling Extremal Events for Insurance and Finance. Berlin: Springer Verlag.

Fama, Eugene F., and Kenneth R. French. 1993, Common Risk Factors in the Returns on Stocksand Bonds, Journal of Financial Economics, 33, 3–56.

Goyal, Amit and Ivo Welch, 2008, A Comprehensive Look at The Empirical Performance of Equity Premium Prediction, Review of Financial Studies, 21, 1455-1508.

Gyuorko, Joseph and Donald B. Keim, 1992, What Does the Stock Market Tell Us About Real Estate Returns?, Journal of the American Real Estate and Urban Economics Association, 20, 457-85.

Gyourko, Joseph, and Edward Nelling, 1996, Systematic risk and diversification in the equity REIT market, Real Estate Economics 24, 493-515.

Hamilton, James D. 1994, Time series analysis, Princeton: Princeton University Press.

Jaffe, Jeffrey, 1991, Taxes and the capital structure of partnerships, REITs, and related entities, Journal of Finance, 46, 401-407.

Lizieri Colin, Satchell Stephen, Zhang Qi, 2007, The underlying return-generating factors for REIT returns: An application of independent component analysis, Real Estate Economics, 35, 569-598.

Mian, Atif R. and Sufi, Amir, 2009, The Consequences of Mortgage Credit Expansion: Evidence from the U.S. Mortgage Default Crisis, Quarterly Journal of Economics, 124(4), 1449-1496.

Muhlhofer, T, 2008, Why do REIT Returns Poorly Reflect Property Returns? Unrealizable Appreciation Gains due to Trading Constraints as the Solution to the Short-Term Disparity, Working Paper, Kelley School of Business, Indiana University.

Ooi, Joseph T. L. Jingliang Wang and James R. Webb, 2009, Idiosyncratic Risk and REIT Returns, The Journal of Real Estate Finance and Economics 38, 420-442

Pavlov and Wachter (2009) REITs and Underlying Real Estate Markets: Is There a Link?, Working Paper.

Peterson, James D. and Cheng-Ho Hsieh, 1997, Do Common Risk Factors in the Returns on Stocks and Bonds Explain Returns on REITs?. Real Estate Economics, 25, 321-345.

Roll, Richard, 2011, The Possible Misdiagnosis of a Crises, Financial Analysts Journal, Forthcoming.

Ross, Stephen A. and Randall C. Zisler 1991, Risk and return in real estate, The Journal of Real Estate Finance and Economics, 4, 175-190.

Scholes, Myron and Joseph T. Williams, 1977, Estimating betas from nonsynchronous data, Journal of Financial Economics 5, 309-327.

Standard and Poors, 2009, S&P/Case-Shiller home price indices – index methodology, November.

Young Michael S. and Richard A. Graff, 1995, Real-Estate is not Normal - A fresh look at Real-Estate Return Distributions, Journal Of Real Estate Finance And Economics, 10, 225-259.

Table 1: Individual REITS

|  | January 1980 | May 2009 |
|---|---|---|
| Total Market Capitalization ($Millions) | $2,453 | $194,993 |

| Number of REITs | | |
|---|---|---|
| All | 90 | 148 |
| Equity | 53 | 115 |
| Mortgage | 29 | 26 |
| Hybrid | 18 | 7 |

| Property Type | | |
|---|---|---|
| Unknown | 8 | 0 |
| Unclassified | 23 | 13 |
| Diversified | 15 | 12 |
| Health Care | 0 | 13 |
| Industrial/Office | 8 | 25 |
| Lodging/Resorts | 3 | 12 |
| Mortgage | 19 | 27 |
| Residential | 8 | 16 |
| Retail | 6 | 25 |
| Self Storage | 0 | 5 |

Table 2: Summary cross-sectional statistics for individual REITs

Annual summary statistics are presented for individual REITs that have a full year of data, 1980 through 2008. The cross-sectional mean and standard deviation (mean of the standard deviations) of return are in percent per month. Betas are computed against the SP500 index.

|  | Number of REITs | Return (%/month) | | Beta | |
|---|---|---|---|---|---|
|  |  | Mean | Standard Deviation | Mean | Standard Deviation |
| 1980 | 80 | 0.764 | 10.876 | 1.044 | 0.611 |
| 1981 | 72 | -0.681 | 9.211 | 0.109 | 0.248 |
| 1982 | 70 | 1.230 | 8.691 | 0.470 | 0.825 |
| 1983 | 70 | 0.742 | 8.884 | 0.536 | 0.910 |
| 1984 | 73 | -0.109 | 6.423 | 0.342 | 0.615 |
| 1985 | 71 | -0.351 | 7.524 | 0.413 | 0.670 |
| 1986 | 91 | 0.183 | 7.655 | 0.325 | 0.514 |
| 1987 | 106 | -2.455 | 9.458 | 0.466 | 0.375 |
| 1988 | 111 | -0.202 | 6.988 | 0.487 | 0.741 |
| 1989 | 120 | -2.268 | 9.941 | 0.471 | 0.698 |
| 1990 | 123 | -4.132 | 13.219 | 0.277 | 0.817 |
| 1991 | 120 | 0.439 | 11.404 | 0.343 | 0.867 |
| 1992 | 137 | -0.679 | 10.754 | -0.060 | 2.270 |
| 1993 | 140 | 1.280 | 11.173 | 0.479 | 2.419 |
| 1994 | 185 | -0.653 | 7.697 | 0.302 | 0.867 |
| 1995 | 220 | 0.765 | 7.228 | 0.227 | 3.507 |
| 1996 | 207 | 1.770 | 6.717 | 0.147 | 1.156 |
| 1997 | 191 | 0.736 | 7.232 | 0.274 | 0.608 |
| 1998 | 200 | -2.238 | 8.512 | 0.489 | 0.700 |
| 1999 | 211 | -1.442 | 8.346 | 0.180 | 0.657 |
| 2000 | 202 | -0.154 | 8.675 | 0.018 | 0.709 |
| 2001 | 192 | 0.829 | 9.268 | 0.296 | 0.680 |
| 2002 | 185 | 0.026 | 7.173 | 0.140 | 0.390 |
| 2003 | 177 | 1.915 | 6.492 | 0.484 | 0.972 |
| 2004 | 173 | 1.260 | 7.887 | 1.129 | 1.259 |
| 2005 | 195 | -0.464 | 6.444 | 1.069 | 0.988 |
| 2006 | 185 | 1.262 | 6.308 | 1.088 | 1.236 |
| 2007 | 161 | -3.238 | 9.606 | 1.515 | 1.331 |
| 2008 | 152 | -6.812 | 20.808 | 1.849 | 1.172 |

Table 3. Descriptive Return Statistics for REIT Indexes and S&P/Case-Shiller® Indexes

REIT results are presented first by REIT type and then by Property type. The mean and standard deviation are in percent per month. Auto-correlation is first order. Beta is against the S&P500. Excess Kurtosis is relative to the normal distribution. Normality is the Jarque-Bera statistic whose 5% critical value is 5.99; hence all series are non-normal. The sample period is January 1980 through May 2009 for REITs and the S&P500 and February 1987 through May 2009 for the SCS indexes.

|  | Mean | Standard Deviation | Skewness | Excess Kurtosis | Normality | Auto correlation | Beta |
|---|---|---|---|---|---|---|---|
| Value-weighted REIT Indexes ||||||||
| All REITs | 0.884 | 4.916 | -0.918 | 11.999 | 1237.3 | 0.138 | 0.625 |
| Equity | 0.933 | 5.062 | -0.898 | 12.974 | 1506.3 | 0.118 | 0.640 |
| Mortgage | 0.527 | 6.317 | -1.027 | 5.794 | 176.4 | 0.103 | 0.631 |
| Hybrid | 0.635 | 5.380 | -0.969 | 7.832 | 397.5 | 0.189 | 0.511 |
| Unclassified | 0.730 | 5.242 | -0.505 | 6.402 | 184.7 | 0.135 | 0.602 |
| Diversified | 0.972 | 5.727 | -0.444 | 12.174 | 1246.0 | 0.148 | 0.642 |
| Industrial Office | 0.717 | 6.387 | -0.639 | 11.504 | 1084.7 | 0.042 | 0.656 |
| Lodging Resort | 0.281 | 8.576 | 0.961 | 14.353 | 1944.7 | 0.164 | 0.824 |
| Residential | 1.005 | 5.549 | -0.731 | 6.836 | 247.2 | 0.084 | 0.636 |
| Retail | 1.077 | 5.672 | -0.546 | 17.321 | 3025.5 | 0.102 | 0.632 |
|  |  |  |  |  |  |  |  |
| S&P500 | 0.594 | 4.539 | -0.962 | 6.353 | 219.2 | 0.076 | 1.000 |

|  | Mean | Standard Deviation | Skewness | Excess Kurtosis | Normality | Auto correlation | Beta |
|---|---|---|---|---|---|---|---|
| S&P/Case-Shiller® (SCS) Indexes ||||||||
| Composite10 | 0.327 | 0.907 | -0.869 | 4.348 | 16.381 | 0.962 | 0.018 |
| Boston | 0.276 | 0.874 | -0.288 | 3.145 | 3.432 | 0.802 | -0.002 |
| Charlotte | 0.235 | 0.544 | -1.043 | 7.256 | 36.381 | 0.673 | 0.018 |
| Chicago | 0.309 | 0.877 | -1.617 | 9.929 | 231.531 | 0.864 | 0.032 |
| Cleveland | 0.227 | 0.748 | -1.945 | 14.443 | 321.325 | 0.539 | 0.030 |
| Denver | 0.333 | 0.694 | -0.912 | 5.058 | 17.876 | 0.805 | 0.019 |
| Las Vegas | 0.197 | 1.425 | -0.461 | 7.833 | 8.145 | 0.938 | 0.028 |
| Los Angeles | 0.370 | 1.302 | -0.528 | 4.007 | 14.335 | 0.954 | 0.027 |
| Miami | 0.283 | 1.180 | -1.312 | 6.061 | 20.654 | 0.931 | 0.017 |
| New York | 0.310 | 0.783 | -0.247 | 3.235 | 14.555 | 0.908 | 0.000 |
| Portland | 0.477 | 0.807 | -0.722 | 6.090 | 20.298 | 0.876 | 0.024 |
| San Diego | 0.364 | 1.238 | -0.293 | 4.553 | 4.729 | 0.928 | 0.023 |
| San Francisco | 0.349 | 1.388 | -0.828 | 5.273 | 9.645 | 0.916 | 0.046 |
| Tampa | 0.223 | 1.020 | -1.043 | 6.541 | 17.548 | 0.894 | 0.034 |
| Washington | 0.359 | 0.997 | -0.338 | 4.065 | 6.301 | 0.931 | 0.018 |

Table 4. Principal Components Analysis of Real Estate Returns.

Principal Components are extracted from the covariance matrix of 6 REIT Indexes and 14 S&P/Case-Shiller® (SCS) residential real estate indexes. Loadings on the first two principal components and R-squares from regressions on the first three principal components for the REIT series and first six principal components for SCS series are reported.

|  | Loading on First PC | Loading on Second PC | Adjusted R-Square |
|---|---|---|---|
|  | Value-weighted REIT Indexes | | |
| Unclassified | 0.299 | 0.208 | 0.903 |
| Diversified | 0.396 | 0.209 | 0.873 |
| Industrial Office | 0.421 | 0.228 | 0.913 |
| Lodging Resort | 0.562 | -0.817 | 0.998 |
| Residential | 0.340 | 0.359 | 0.793 |
| Retail | 0.381 | 0.253 | 0.854 |

|  | S&P/Case-Shiller® (SCS) Indexes | | |
|---|---|---|---|
| Boston | 0.189 | -0.396 | 0.918 |
| Charlotte | 0.090 | -0.112 | 0.566 |
| Chicago | 0.208 | -0.103 | 0.742 |
| Cleveland | 0.142 | -0.263 | 0.772 |
| Denver | 0.109 | -0.244 | 0.775 |
| Las Vegas | 0.376 | 0.578 | 0.983 |
| Los Angeles | 0.380 | -0.053 | 0.933 |
| Miami | 0.321 | 0.370 | 0.943 |
| New York | 0.189 | -0.074 | 0.892 |
| Portland | 0.160 | 0.135 | 0.833 |
| San Diego | 0.351 | -0.125 | 0.941 |
| San Francisco | 0.388 | -0.359 | 0.925 |
| Tampa | 0.272 | 0.204 | 0.915 |
| Washington | 0.289 | -0.087 | 0.877 |

Table 5. Measures of the Risk of Loss for Real Estate Index Returns

Risk of loss estimates are reported for three different models of the distribution of real estate index returns: (1) EM, the Efficient Maximization algorithm fit to the unconditional distribution; (2) The Generalized Pareto Distribution (GPD); and (3) GARCH (1,1). "Loss" is the <u>negative</u> of the loss distribution's fractile for the probability reported, which indicates the likelihood that the monthly observed return will exceed minus this level. "Average loss" is the negative of the expected return conditional on the loss exceeding the reported loss fractile.

|  | Loss Fractile | EM | | GPD | | GARCH | |
|---|---|---|---|---|---|---|---|
|  |  | Loss | Average Loss | Loss | Average Loss | Loss | Average Loss |
| Value-weighted REIT Indexes | | | | | | | |
| All REITs | 0.95 | 5.689 | 8.391 | 7.539 | 10.613 | 6.565 | 9.639 |
|  | 0.99 | 8.461 | 10.842 | 12.269 | 16.283 | 11.296 | 15.310 |
|  | 0.999 | 11.569 | 13.697 | 21.658 | 27.539 | 20.685 | 26.566 |
| Equity | 0.95 | 5.877 | 8.540 | 7.515 | 10.895 | 6.480 | 9.860 |
|  | 0.99 | 8.699 | 11.034 | 12.552 | 17.560 | 11.517 | 16.525 |
|  | 0.999 | 11.861 | 13.940 | 24.250 | 33.036 | 23.215 | 32.001 |
| Mortgage | 0.95 | 8.272 | 11.035 | 9.079 | 12.150 | 8.379 | 11.450 |
|  | 0.99 | 11.918 | 14.258 | 14.006 | 17.154 | 13.307 | 16.454 |
|  | 0.999 | 16.004 | 18.013 | 21.270 | 24.531 | 20.570 | 23.831 |
| Hybrid | 0.95 | 7.184 | 9.806 | 8.213 | 11.320 | 7.300 | 10.407 |
|  | 0.99 | 10.424 | 12.670 | 13.043 | 16.910 | 12.130 | 15.996 |
|  | 0.999 | 14.055 | 16.007 | 22.076 | 27.365 | 21.163 | 26.452 |
| Unclassified | 0.95 | 7.073 | 9.785 | 8.739 | 11.818 | 7.956 | 11.035 |
|  | 0.99 | 10.306 | 12.643 | 13.705 | 16.732 | 12.921 | 15.949 |
|  | 0.999 | 13.929 | 15.973 | 20.665 | 23.621 | 19.882 | 22.838 |
| Diversified | 0.95 | 6.829 | 9.783 | 8.708 | 12.782 | 7.732 | 11.807 |
|  | 0.99 | 10.061 | 12.641 | 14.988 | 20.269 | 14.013 | 19.293 |
|  | 0.999 | 13.684 | 15.970 | 27.338 | 34.990 | 26.362 | 34.014 |
| Industrial Office | 0.95 | 7.794 | 10.956 | 9.157 | 14.955 | 8.219 | 14.017 |
|  | 0.99 | 11.414 | 14.156 | 17.416 | 27.251 | 16.478 | 26.312 |
|  | 0.999 | 15.471 | 17.884 | 40.146 | 61.091 | 39.208 | 60.153 |
| Lodging Resort | 0.95 | 11.715 | 15.043 | 11.607 | 19.807 | 10.655 | 18.855 |
|  | 0.99 | 16.685 | 19.437 | 23.469 | 36.801 | 22.517 | 35.849 |
|  | 0.999 | 22.256 | 24.556 | 54.421 | 81.145 | 53.469 | 80.193 |
| Residential | 0.95 | 7.117 | 10.185 | 9.634 | 12.590 | 8.679 | 11.634 |
|  | 0.99 | 10.482 | 13.160 | 14.448 | 17.083 | 13.492 | 16.128 |
|  | 0.999 | 14.254 | 16.626 | 20.444 | 22.682 | 19.488 | 21.726 |
| Retail | 0.95 | 6.314 | 9.268 | 8.205 | 12.392 | 6.934 | 11.121 |
|  | 0.99 | 9.376 | 11.975 | 14.355 | 20.860 | 13.084 | 19.589 |
|  | 0.999 | 12.808 | 15.128 | 29.512 | 41.727 | 28.241 | 40.456 |
| SP500 | 0.95 | 6.009 | 8.280 | 7.315 | 8.950 | 6.629 | 8.265 |
|  | 0.99 | 8.744 | 10.698 | 10.023 | 11.134 | 9.337 | 10.448 |
|  | 0.999 | 11.810 | 13.515 | 12.460 | 13.100 | 11.775 | 12.414 |

| \multicolumn{8}{c}{S&P/Case-Shiller® (SCS) Indexes} |
|---|---|---|---|---|---|---|---|
| Composite10 | 0.95 | 0.826 | 1.446 | 1.625 | 1.868 | 1.071 | 1.314 |
| | 0.99 | 1.303 | 1.868 | 2.028 | 2.165 | 1.474 | 1.611 |
| | 0.999 | 1.839 | 2.360 | 2.319 | 2.379 | 1.765 | 1.825 |
| Boston | 0.95 | 1.011 | 1.614 | 1.507 | 1.954 | 1.196 | 1.643 |
| | 0.99 | 1.544 | 2.085 | 2.173 | 2.833 | 1.863 | 2.522 |
| | 0.999 | 2.141 | 2.634 | 3.714 | 4.864 | 3.403 | 4.553 |
| Charlotte | 0.95 | 0.543 | 0.976 | 1.027 | 1.289 | 0.771 | 1.032 |
| | 0.99 | 0.866 | 1.261 | 1.448 | 1.707 | 1.192 | 1.450 |
| | 0.999 | 1.227 | 1.593 | 2.042 | 2.296 | 1.786 | 2.040 |
| Chicago | 0.95 | 0.862 | 1.469 | 1.528 | 1.933 | 1.105 | 1.510 |
| | 0.99 | 1.348 | 1.898 | 2.193 | 2.513 | 1.770 | 2.090 |
| | 0.999 | 1.892 | 2.398 | 2.910 | 3.139 | 2.487 | 2.715 |
| Cleveland | 0.95 | 0.784 | 1.268 | 1.183 | 1.544 | 0.847 | 1.208 |
| | 0.99 | 1.203 | 1.638 | 1.756 | 2.156 | 1.420 | 1.820 |
| | 0.999 | 1.672 | 2.069 | 2.685 | 3.148 | 2.349 | 2.813 |
| Denver | 0.95 | 0.636 | 1.215 | 1.325 | 1.569 | 0.957 | 1.201 |
| | 0.99 | 1.037 | 1.570 | 1.727 | 1.910 | 1.359 | 1.541 |
| | 0.999 | 1.487 | 1.983 | 2.133 | 2.254 | 1.765 | 1.886 |
| Las Vegas | 0.95 | 1.536 | 2.173 | 1.829 | 4.062 | 1.390 | 2.974 |
| | 0.99 | 2.254 | 2.808 | 4.288 | 10.059 | 4.733 | 5.042 |
| | 0.999 | 3.058 | 3.547 | 15.880 | 38.326 | 5.321 | 5.494 |
| Los Angeles | 0.95 | 1.578 | 2.442 | 2.223 | 2.797 | 1.690 | 2.264 |
| | 0.99 | 2.384 | 3.156 | 3.126 | 3.796 | 2.593 | 3.264 |
| | 0.999 | 3.289 | 3.987 | 4.687 | 5.524 | 4.155 | 4.992 |
| Miami | 0.95 | 0.741 | 1.284 | 1.987 | 2.322 | 1.664 | 1.999 |
| | 0.99 | 0.887 | 1.660 | 2.542 | 2.691 | 2.220 | 2.369 |
| | 0.999 | 1.641 | 2.097 | 2.848 | 2.895 | 2.525 | 2.572 |
| New York | 0.95 | 0.916 | 1.537 | 1.509 | 1.782 | 1.314 | 1.588 |
| | 0.99 | 1.423 | 1.986 | 1.956 | 2.186 | 1.762 | 1.992 |
| | 0.999 | 1.993 | 2.509 | 2.477 | 2.657 | 2.282 | 2.462 |
| Portland | 0.95 | 0.685 | 1.458 | 1.720 | 2.154 | 1.182 | 1.615 |
| | 0.99 | 1.166 | 1.883 | 2.432 | 2.777 | 1.894 | 2.238 |
| | 0.999 | 1.706 | 2.379 | 3.205 | 3.453 | 2.666 | 2.914 |
| San Diego | 0.95 | 1.239 | 2.010 | 2.147 | 2.915 | 1.506 | 2.273 |
| | 0.99 | 1.765 | 2.597 | 3.322 | 4.347 | 2.681 | 3.706 |
| | 0.999 | 2.647 | 3.280 | 5.720 | 7.270 | 5.078 | 6.629 |
| San Francisco | 0.95 | 1.635 | 2.488 | 2.435 | 3.013 | 1.950 | 2.528 |
| | 0.99 | 2.457 | 3.215 | 3.391 | 3.799 | 2.906 | 3.314 |
| | 0.999 | 3.378 | 4.062 | 4.291 | 4.539 | 3.806 | 4.053 |
| Tampa | 0.95 | 0.939 | 1.458 | 1.791 | 2.214 | 1.539 | 1.962 |
| | 0.99 | 1.038 | 1.883 | 2.493 | 2.755 | 2.241 | 2.502 |
| | 0.999 | 1.960 | 2.379 | 3.058 | 3.189 | 2.806 | 2.937 |
| Washington | 0.95 | 0.810 | 1.466 | 1.915 | 2.369 | 1.745 | 2.199 |
| | 0.99 | 1.209 | 1.894 | 2.657 | 3.040 | 2.487 | 2.870 |
| | 0.999 | 1.837 | 2.393 | 3.521 | 3.820 | 3.351 | 3.650 |

Table 6. Measures of the Risk of Loss for Real Estate Index Residuals

Risk of loss estimates are reported for three different models of the distribution of real estate index residuals: (1) EM, the Efficient Maximization algorithm fit to the unconditional distribution; (2) The Generalized Pareto Distribution (GPD); and (3) GARCH (1,1). "Loss" is the <u>negative</u> of the loss distribution's fractile for the probability reported, which indicates the likelihood that the monthly observed return will exceed minus this level. "Average loss" is the negative of the expected return given conditional on the loss exceeding the reported loss fractile. The residuals are obtained by fitting returns to the first three principal components for the REIT series and first six principal components for SCS series.

|  | Loss Fractile | EM Loss | EM Average Loss | GPD Loss | GPD Average Loss | GARCH Loss | GARCH Average Loss |
|---|---|---|---|---|---|---|---|
| Value-weighted REIT Indexes ||||||||
| All REITs | 0.95 | 1.331 | 1.669 | 1.503 | 1.866 | 1.511 | 1.874 |
|  | 0.99 | 1.883 | 2.157 | 2.106 | 2.299 | 2.115 | 2.308 |
|  | 0.999 | 2.501 | 2.725 | 2.515 | 2.593 | 2.523 | 2.602 |
| Equity | 0.95 | 1.205 | 1.511 | 1.200 | 1.641 | 1.204 | 1.646 |
|  | 0.99 | 1.704 | 1.952 | 1.910 | 2.353 | 1.914 | 2.358 |
|  | 0.999 | 2.264 | 2.466 | 2.931 | 3.377 | 2.936 | 3.381 |
| Mortgage | 0.95 | 2.537 | 3.165 | 3.027 | 4.240 | 2.992 | 4.205 |
|  | 0.99 | 3.583 | 4.090 | 5.013 | 6.028 | 4.978 | 5.993 |
|  | 0.999 | 4.755 | 5.167 | 7.304 | 8.091 | 7.269 | 8.056 |
| Hybrid | 0.95 | 3.029 | 3.798 | 3.038 | 3.903 | 3.043 | 3.908 |
|  | 0.99 | 4.284 | 4.908 | 4.457 | 5.158 | 4.462 | 5.162 |
|  | 0.999 | 5.690 | 6.200 | 6.032 | 6.551 | 6.037 | 6.556 |
| Unclassified | 0.95 | 3.249 | 4.075 | 3.462 | 5.531 | 3.463 | 5.604 |
|  | 0.99 | 4.596 | 5.265 | 6.654 | 9.325 | 6.816 | 8.068 |
|  | 0.999 | 6.105 | 6.652 | 12.901 | 16.750 | 9.095 | 9.449 |
| Diversified | 0.95 | 2.846 | 3.569 | 3.174 | 4.060 | 3.180 | 4.065 |
|  | 0.99 | 4.025 | 4.612 | 4.638 | 5.267 | 4.644 | 5.273 |
|  | 0.999 | 5.347 | 5.826 | 6.028 | 6.414 | 6.033 | 6.420 |
| Industrial Office | 0.95 | 2.117 | 2.655 | 2.062 | 2.919 | 1.999 | 2.856 |
|  | 0.99 | 2.994 | 3.431 | 3.407 | 4.419 | 3.345 | 4.357 |
|  | 0.999 | 3.978 | 4.334 | 5.767 | 7.051 | 5.704 | 6.988 |
| Lodging Resort | 0.95 | 2.577 | 3.231 | 2.747 | 4.268 | 2.713 | 4.234 |
|  | 0.99 | 3.644 | 4.175 | 5.135 | 6.931 | 5.101 | 6.897 |
|  | 0.999 | 4.841 | 5.275 | 9.320 | 11.597 | 9.286 | 11.563 |
| Residential | 0.95 | 1.792 | 2.247 | 1.696 | 2.505 | 1.796 | 2.605 |
|  | 0.99 | 2.534 | 2.903 | 2.927 | 4.031 | 3.028 | 4.131 |
|  | 0.999 | 3.367 | 3.668 | 5.509 | 7.230 | 5.609 | 7.330 |
| Retail | 0.95 | 2.100 | 2.633 | 2.112 | 2.671 | 2.083 | 2.642 |
|  | 0.99 | 2.969 | 3.402 | 3.034 | 3.443 | 3.005 | 3.415 |
|  | 0.999 | 3.944 | 4.298 | 3.942 | 4.204 | 3.913 | 4.175 |
| SP500 | 0.95 | 2.894 | 3.629 | 3.079 | 4.439 | 3.126 | 4.516 |
|  | 0.99 | 4.093 | 4.690 | 5.221 | 6.798 | 4.924 | 6.270 |
|  | 0.999 | 5.437 | 5.924 | 8.894 | 10.845 | 7.524 | 8.450 |
| S&P/Case-Shiller® (SCS) Indexes ||||||||
| Composite10 | 0.95 | 0.111 | 0.138 | 1.625 | 1.868 | 1.071 | 1.314 |
|  | 0.99 | 0.156 | 0.179 | 2.028 | 2.165 | 1.474 | 1.611 |
|  | 0.999 | 0.207 | 0.226 | 2.319 | 2.379 | 1.765 | 1.825 |
| Boston | 0.95 | 0.372 | 0.466 | 1.507 | 1.954 | 1.196 | 1.643 |
|  | 0.99 | 0.526 | 0.602 | 2.173 | 2.833 | 1.863 | 2.522 |
|  | 0.999 | 0.698 | 0.760 | 3.714 | 4.864 | 3.403 | 4.553 |
| Charlotte | 0.95 | 0.607 | 0.762 | 1.027 | 1.289 | 0.771 | 1.032 |
|  | 0.99 | 0.859 | 0.984 | 1.448 | 1.707 | 1.192 | 1.450 |
|  | 0.999 | 1.141 | 1.244 | 2.042 | 2.296 | 1.786 | 2.040 |
| Chicago | 0.95 | 0.581 | 0.728 | 1.528 | 1.933 | 1.105 | 1.510 |
|  | 0.99 | 0.821 | 0.941 | 2.193 | 2.513 | 1.770 | 2.090 |
|  | 0.999 | 1.091 | 1.189 | 2.910 | 3.139 | 2.487 | 2.715 |

| City | | | | | | | |
|---|---|---|---|---|---|---|---|
| Cleveland | 0.95 | 0.528 | 0.662 | 1.183 | 1.544 | 0.847 | 1.208 |
| | 0.99 | 0.747 | 0.855 | 1.756 | 2.156 | 1.420 | 1.820 |
| | 0.999 | 0.992 | 1.081 | 2.685 | 3.148 | 2.349 | 2.813 |
| Denver | 0.95 | 0.453 | 0.568 | 1.325 | 1.569 | 0.957 | 1.201 |
| | 0.99 | 0.641 | 0.734 | 1.727 | 1.910 | 1.359 | 1.541 |
| | 0.999 | 0.851 | 0.928 | 2.133 | 2.254 | 1.765 | 1.886 |
| Las Vegas | 0.95 | 0.263 | 0.330 | 0.289 | 0.387 | 0.278 | 0.411 |
| | 0.99 | 0.372 | 0.426 | 0.449 | 0.533 | 0.457 | 0.544 |
| | 0.999 | 0.494 | 0.538 | 0.639 | 0.706 | 0.591 | 0.618 |
| Los Angeles | 0.95 | 0.416 | 0.520 | 0.419 | 0.546 | 0.427 | 0.554 |
| | 0.99 | 0.587 | 0.671 | 0.625 | 0.746 | 0.634 | 0.754 |
| | 0.999 | 0.780 | 0.848 | 0.901 | 1.012 | 0.909 | 1.020 |
| Miami | 0.95 | 0.412 | 0.518 | 1.987 | 2.322 | 1.664 | 1.999 |
| | 0.99 | 0.583 | 0.669 | 2.542 | 2.691 | 2.220 | 2.369 |
| | 0.999 | 0.775 | 0.845 | 2.848 | 2.895 | 2.525 | 2.572 |
| New York | 0.95 | 0.400 | 0.500 | 1.509 | 1.782 | 1.314 | 1.588 |
| | 0.99 | 0.565 | 0.647 | 1.956 | 2.186 | 1.762 | 1.992 |
| | 0.999 | 0.750 | 0.817 | 2.477 | 2.657 | 2.282 | 2.462 |
| Portland | 0.95 | 0.580 | 0.727 | 1.720 | 2.154 | 1.182 | 1.615 |
| | 0.99 | 0.820 | 0.939 | 2.432 | 2.777 | 1.894 | 2.238 |
| | 0.999 | 1.089 | 1.186 | 3.205 | 3.453 | 2.666 | 2.914 |
| San Diego | 0.95 | 0.481 | 0.604 | 2.147 | 2.915 | 1.506 | 2.273 |
| | 0.99 | 0.680 | 0.780 | 3.322 | 4.347 | 2.681 | 3.706 |
| | 0.999 | 0.904 | 0.985 | 5.720 | 7.270 | 5.078 | 6.629 |
| San Francisco | 0.95 | 0.323 | 0.404 | 2.435 | 3.013 | 1.950 | 2.528 |
| | 0.99 | 0.456 | 0.522 | 3.391 | 3.799 | 2.906 | 3.314 |
| | 0.999 | 0.606 | 0.660 | 4.291 | 4.539 | 3.806 | 4.053 |
| Tampa | 0.95 | 0.450 | 0.564 | 1.791 | 2.214 | 1.539 | 1.962 |
| | 0.99 | 0.637 | 0.728 | 2.493 | 2.755 | 2.241 | 2.502 |
| | 0.999 | 0.845 | 0.920 | 3.058 | 3.189 | 2.806 | 2.937 |
| Washington | 0.95 | 0.515 | 0.647 | 1.915 | 2.369 | 1.745 | 2.199 |
| | 0.99 | 0.729 | 0.837 | 2.657 | 3.040 | 2.487 | 2.870 |
| | 0.999 | 0.968 | 1.057 | 3.521 | 3.820 | 3.351 | 3.650 |

Table 7. Prediction of Real Estate Index Returns

Predictions for a sample of series are presented: All REITs Value Weighted Index, REITs Residential Property Value Weighted Index respectively with the SCS Composite-10 index. A MA(3) of both REIT series is used. The unit root tests are reported in Panel A. ADF is the Augmented Dickey Fuller test, PP is the Phillips-Perron test and Statitionarity is the KPSS test. Each unit root test statistic is followed by p-values. The symbol * indicates significance at the 5% level and the symbol ** indicates significance at the 1% level. The VAR model coefficients are reported in Panel B. For the VAR, the optimal lag order is 5 based on both the AIC and BIC criterion. VAR Model coefficients are followed by respective t-statistics. Granger causality tests are reported in Panel C with test statistics followed by p-values. The VAR model forecasts are reported in Panel D for 6 periods with coefficients followed by standard errors.

| Panel A: Unit Root Tests | | | |
|---|---|---|---|
| | ADF | PP | Stationarity |
| Log Prices | | | |
| Residential | -0.412 | 0.838 | 0.765** |
| | (-0.987) | (-1.000) | |
| All REITs | -0.258 | 1.234 | 0.589** |
| | (-0.991) | (-1.000) | |
| Composite | -2.036 | -0.35 | 0.794** |
| | (-0.578) | (-0.989) | |
| | | | |
| Log Returns | | | |
| Residential | -4.374 | -5.285 | 0.610* |
| | (0.000) | (0.000) | |
| All REITs | -4.822 | -5.736 | 0.594* |
| | (0.000) | (0.000) | |
| Composite | -2.746 | -3.532 | 0.45 |
| | (-0.068) | (-0.008) | |

| Panel B: VAR Model Coefficients | | | | | |
|---|---|---|---|---|---|
| | Composite10 | Residential | | Composite10 | All REITs |
| Intercept | 0.004 | 0.018 | Intercept | 0.005 | 0.016 |
| | (1.110) | (1.624) | | (1.267) | (2.013) |
| Composite10 Lag1 | 1.268 | -0.056 | Composite10 Lag1 | 1.232 | 0.118 |
| | (20.392) | (-0.282) | | (19.909) | (0.825) |
| Residential Lag1 | 0.007 | 1.105 | All REITs Lag1 | 0.035 | 1.011 |
| | (0.359) | (17.115) | | (1.179) | (14.630) |

| | Composite10 | Residential | | Composite10 | All REITs |
|---|---|---|---|---|---|
| Composite10 Lag2 | -0.033 | 0.614 | Composite10 Lag2 | 0.006 | 0.121 |
| | (-0.319) | (1.890) | | (0.057) | (0.525) |
| Residential Lag2 | 0.052 | -0.247 | All REITs Lag2 | 0.078 | -0.140 |
| | (1.902) | (-2.819) | | (1.891) | (-1.466) |
| Composite10 Lag3 | -0.400 | -0.713 | Composite10 Lag3 | -0.399 | -0.147 |
| | (-3.856) | (-2.156) | | (-3.980) | (-0.636) |
| Residential Lag3 | -0.068 | -0.447 | All REITs Lag3 | -0.094 | -0.445 |
| | (-2.644) | (-5.486) | | (-2.457) | (-5.037) |
| Composite10 Lag4 | -0.154 | 0.235 | Composite10 Lag4 | -0.152 | -0.151 |
| | (-1.423) | (0.679) | | (-1.442) | (-0.622) |
| Residential Lag4 | -0.005 | 0.627 | All REITs Lag4 | -0.019 | 0.565 |
| | (-0.196) | (7.332) | | (-0.481) | (6.094) |
| Composite10 Lag5 | 0.262 | -0.100 | Composite10 Lag5 | 0.257 | 0.092 |
| | (3.982) | (-0.477) | | (3.998) | (0.623) |
| Residential Lag5 | 0.011 | -0.239 | All REITs Lag5 | -0.011 | -0.275 |
| | (0.521) | (-3.640) | | (-0.346) | (-3.714) |
| R-squared | 0.927 | 0.741 | R-squared | 0.927 | 0.684 |
| Adjusted R-squared | 0.924 | 0.730 | Adjusted R-squared | 0.924 | 0.671 |

| Panel C: Granger Causality | | | | | |
|---|---|---|---|---|---|
| | Composite10 | Residential | | Composite10 | All REITs |
| Composite10 | | 3.354 | Composite10 | | 5.569 |
| | | (0.006) | | | (0.000) |
| Residential | 3.191 | | All REITs | 8.673 | |
| | (0.008) | | | (0.000) | |

| Panel D: VAR Model Forecasts | | | | | |
|---|---|---|---|---|---|
| | Composite10 | Residential | | Composite10 | All REITs |
| 1-step-ahead | 0.230 | 0.602 | 1-step-ahead | 0.225 | 0.530 |
| | (0.051) | (0.162) | | (0.050) | (0.115) |
| 2-step-ahead | 0.349 | 0.654 | 2-step-ahead | 0.399 | 0.759 |
| | (0.082) | (0.240) | | (0.079) | (0.164) |
| 3-step-ahead | 0.345 | 0.431 | 3-step-ahead | 0.444 | 0.434 |
| | (0.116) | (0.290) | | (0.113) | (0.195) |
| 4-step-ahead | 0.280 | 0.317 | 4-step-ahead | 0.446 | 0.445 |
| | (0.142) | (0.298) | | (0.139) | (0.201) |

| | 0.172 | 0.330 | | 0.350 | 0.333 |
| --- | --- | --- | --- | --- | --- |
| 5-step-ahead | (0.158) | (0.306) | 5-step-ahead | (0.156) | (0.207) |
| | 0.073 | 0.354 | | 0.243 | 0.371 |
| 6-step-ahead | (0.166) | (0.310) | 6-step-ahead | (0.165) | (0.210) |

Figure 1. Time Series plots of Returns Series

Monthly returns for a sample of series are presented: All REITs Value Weighted Index, REITs Residential Property Value Weighted Index and SCS Composite-10 index. The sample period is January 1980 through May 2009 for REITs and February 1987 through May 2009 for the SCS index.

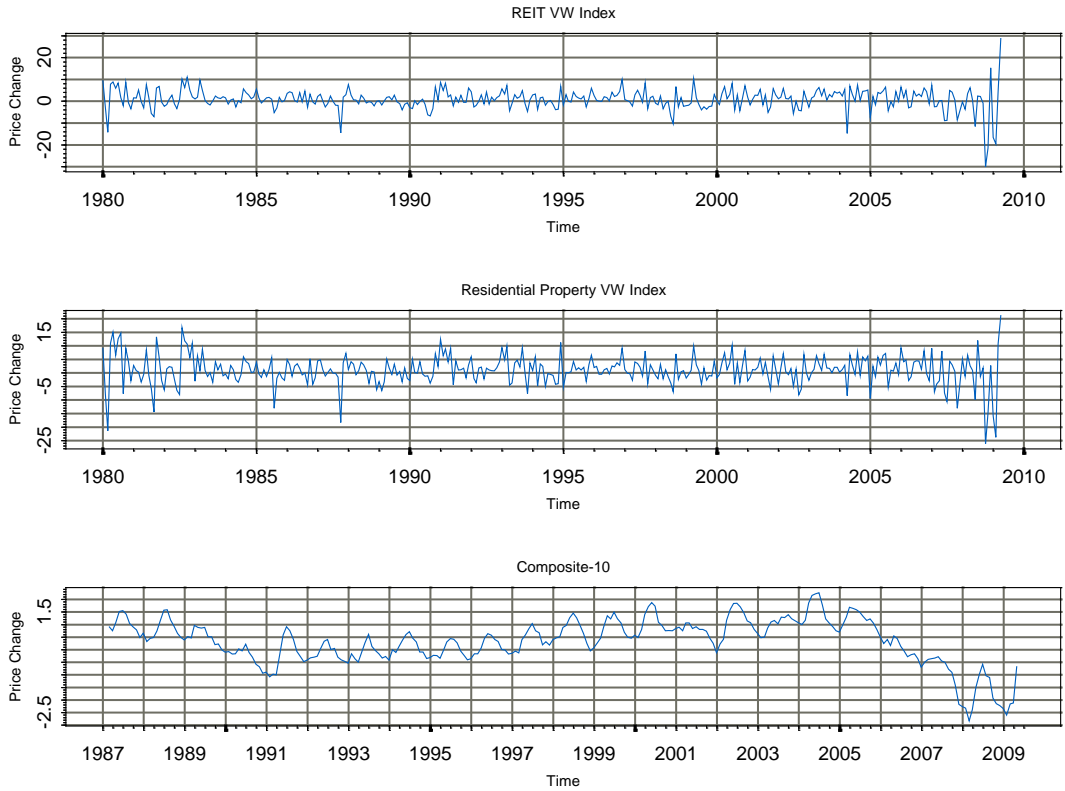

Figure 2. Time Series plots of Conditional Volatility Series

Monthly conditional volatility (from fitting a GARCH (1, 1) model) for a sample of series are presented: All REITs Value Weighted Index, REITs Residential Property Value Weighted Index and SCS Composite-10 index. The sample period is January 1980 through May 2009 for REITs and February 1987 through May 2009 for the SCS index.

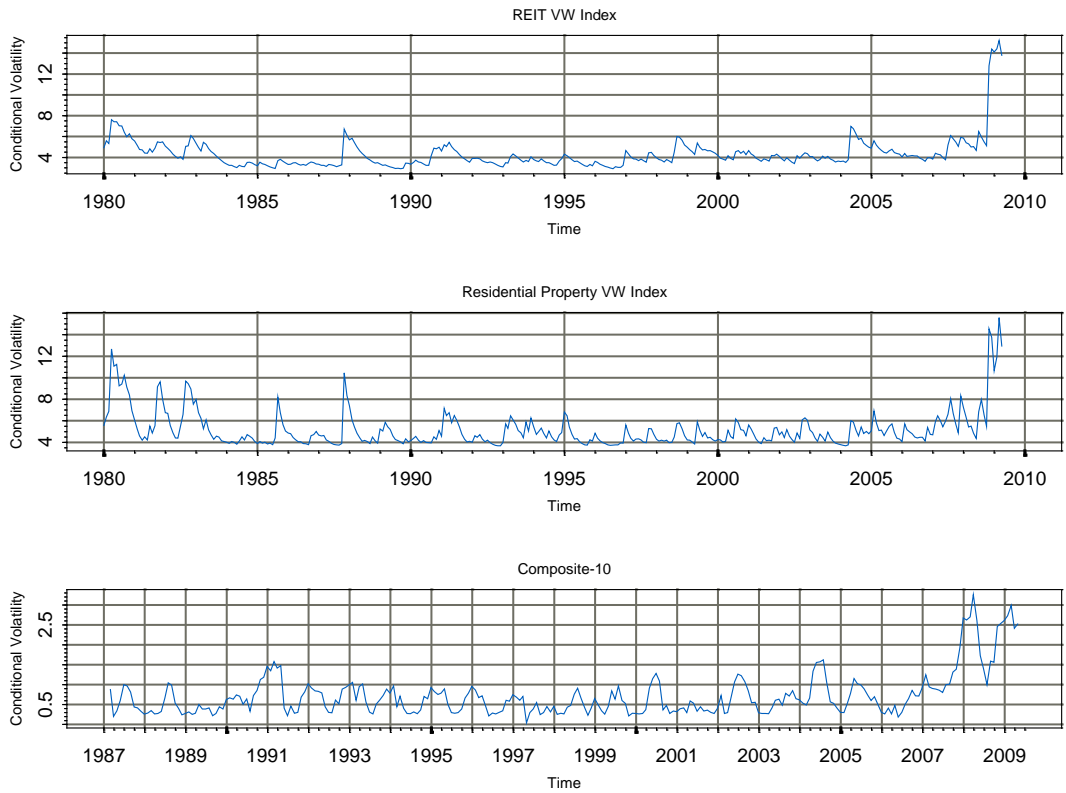

Figure 3. Scree plot for REIT Indexes

For the six REIT indexes whose descriptive statistics are reported in Table 1, excluding the aggregated series, a covariance matrix is formed using their concurrent monthly returns, January 1980 through May 2009. The scree plots depict the variance explained by successive principal components (PCs) with PCs ranked from left to right in descending order of variance explained. The bar gives the variance explained by the PC and the cumulative variance explained in percent is printed above each bar.

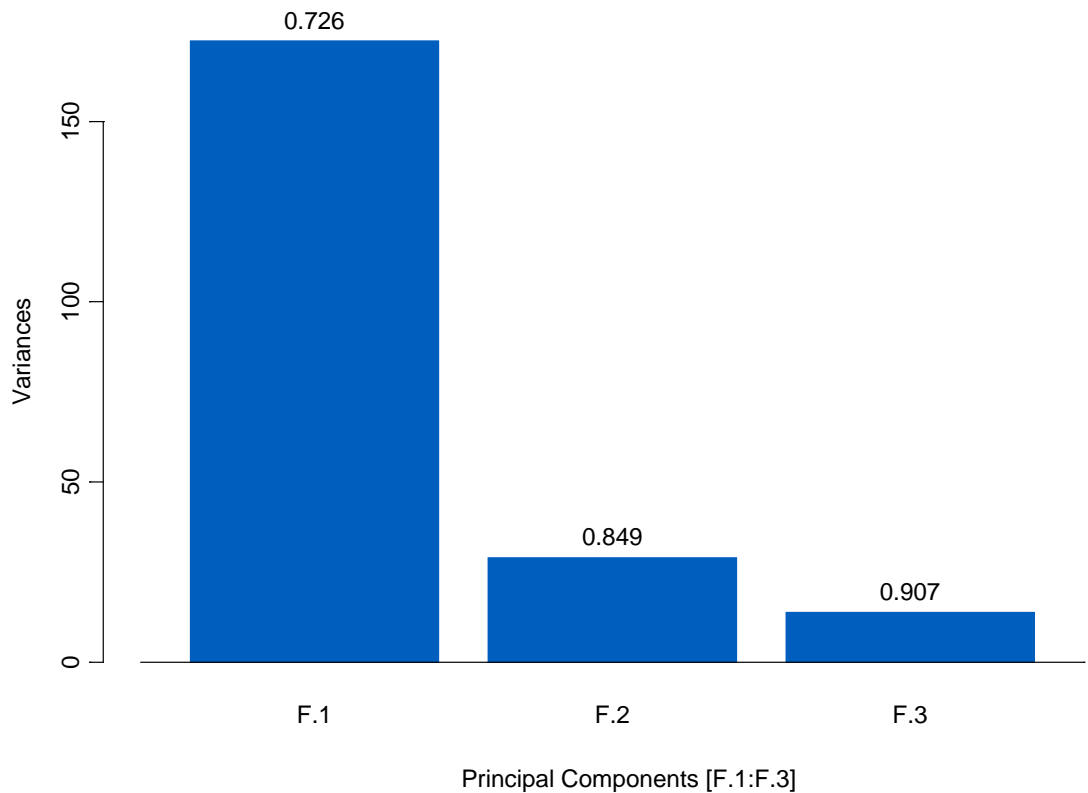

Figure 4. Scree plot for S&P/Case-Shiller® (SCS) Indexes

For the 14 SCS indexes whose descriptive statistics are reported in Table 1 excluding the composite series, a covariance matrix is formed using their concurrent monthly returns, February 1987 through May 2009. The scree plots depict the variance explained by successive principal components (PCs) with PCs ranked from left to right in descending order of variance explained. The bar gives the variance explained by the PC and the cumulative variance explained in percent is printed above each bar.

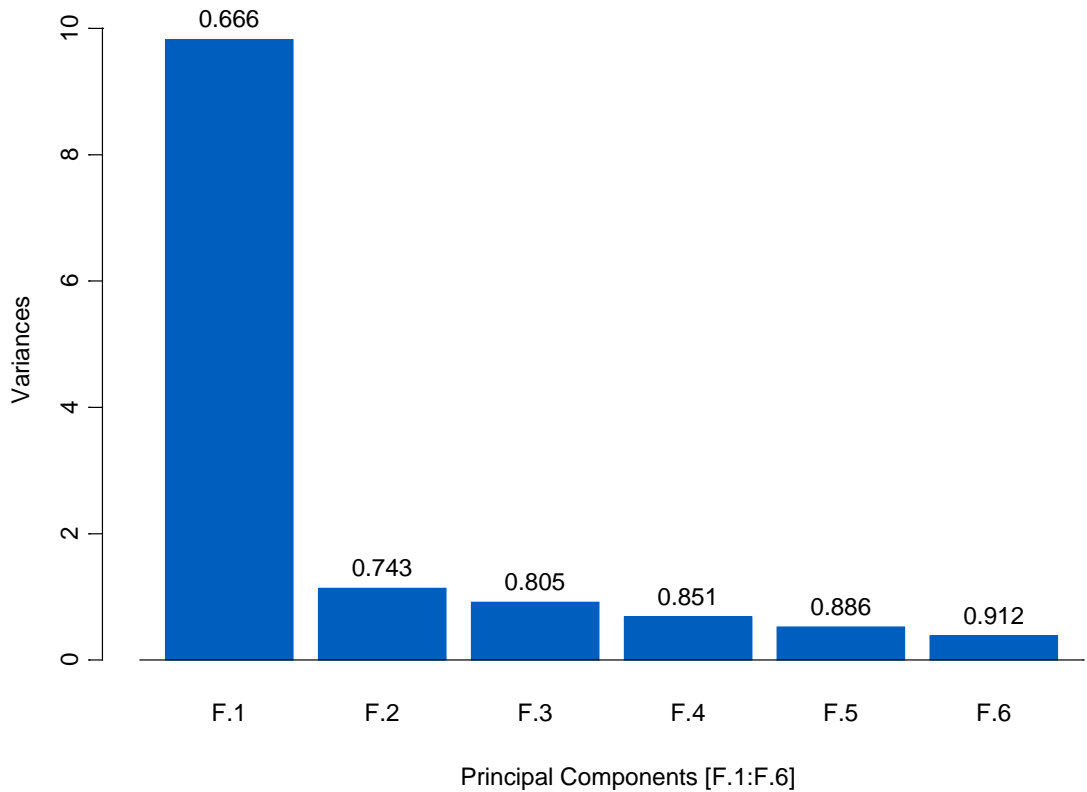

Figure 5. Time Series plots of Log Price Series

Monthly log prices for a sample of series are presented: All REITs Value Weighted Index, REITs Residential Property Value Weighted Index and SCS Composite-10 index. The sample period is February 1987 through May 2009 for the three series.

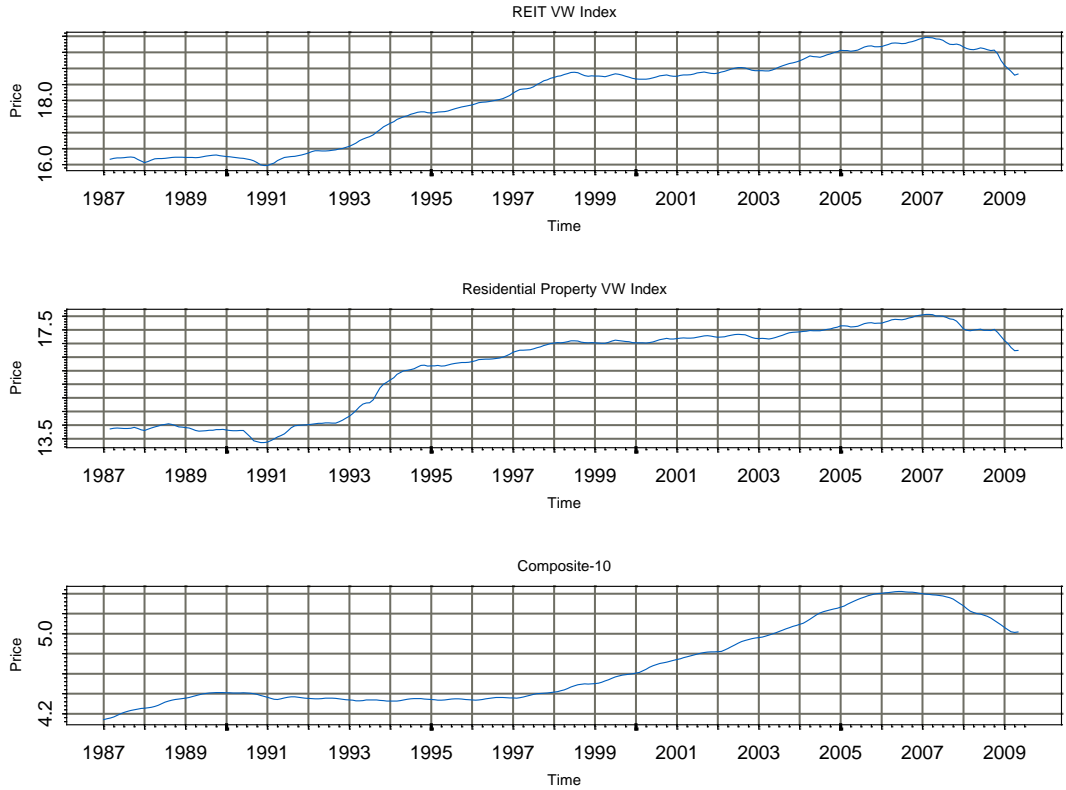

Figure 6. Time Series plots of Response and Fitted Values

Monthly response and fitted values from the VAR model for a sample of series are presented: All REITs Value Weighted Index, REITs Residential Property Value Weighted Index respectively with the SCS Composite-10 index. The sample period is February 1987 through May 2009 for the three series.

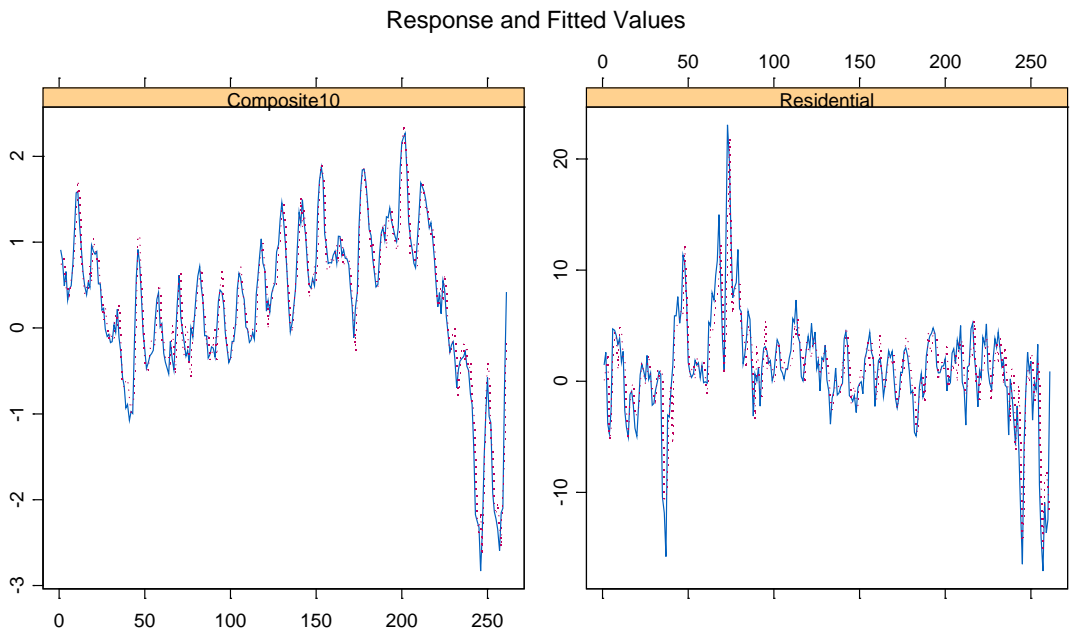

Figure 7. Autocorrelation and cross correlation of Real Estate Index Returns

Autocorrelation and cross correlation with confidence bands for a sample of series are presented: All REITs Value Weighted Index, REITs Residential Property Value Weighted Index respectively with the SCS Composite-10 index. The sample period is February 1987 through May 2009 for the three series.

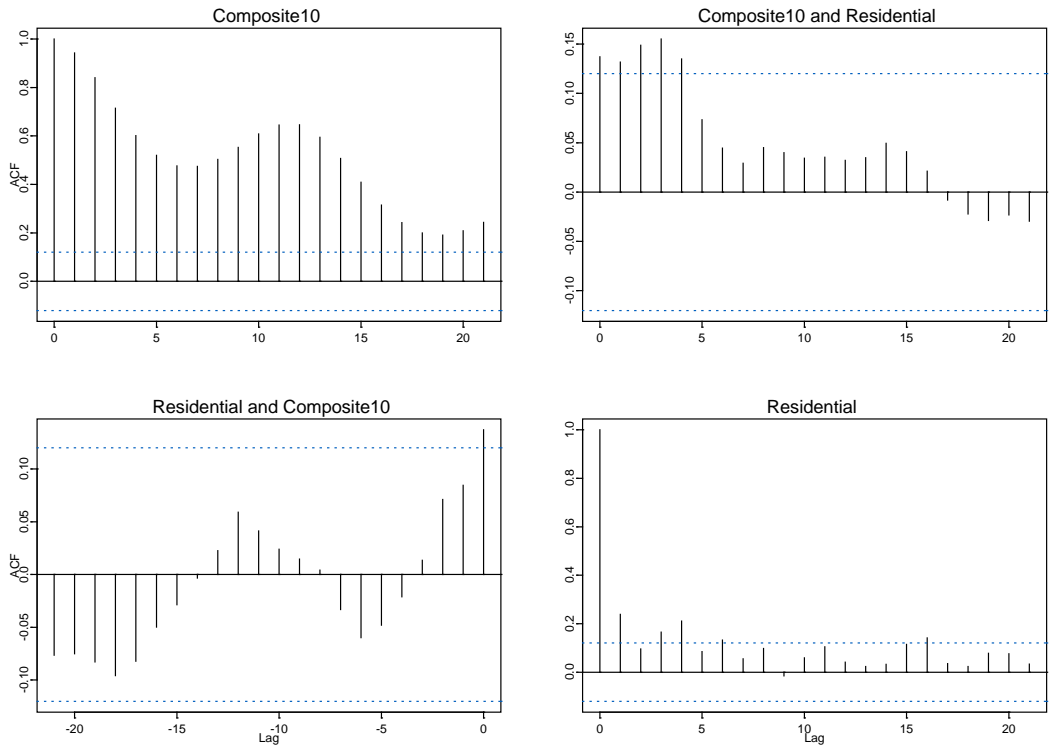

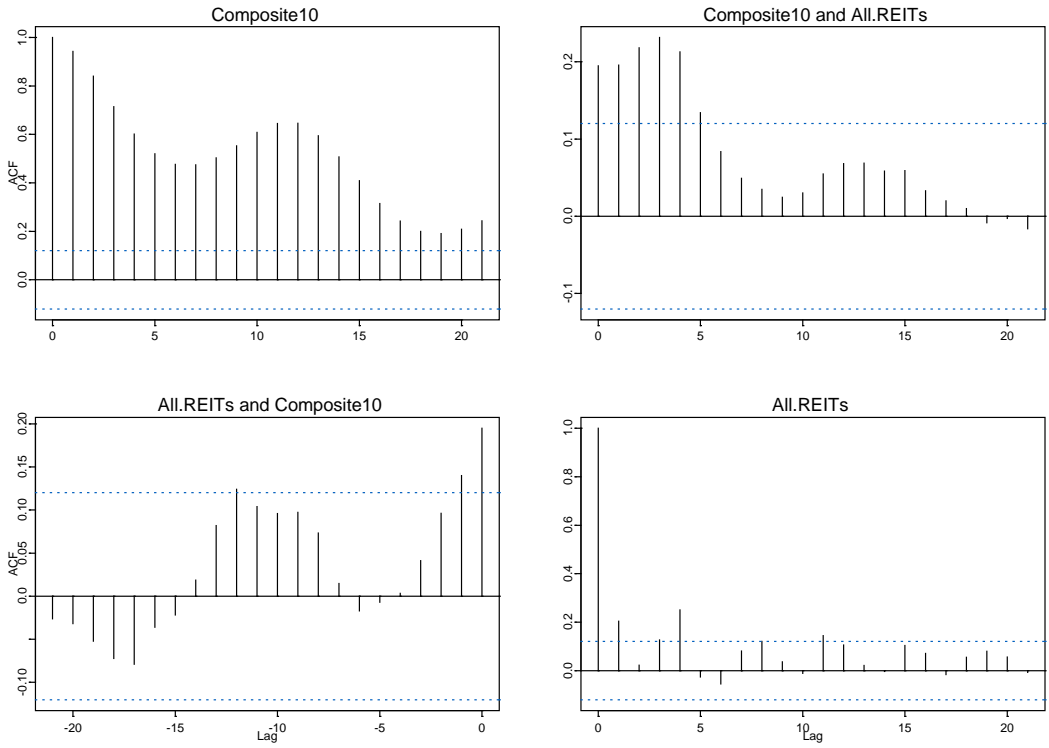

Figure 8. Impulse Response Functions for Real Estate Returns

Impulse response functions and asymptotic standard errors from the VAR model for a sample of series are presented: All REITs Value Weighted Index, REITs Residential Property Value Weighted Index respectively with the SCS Composite-10 index. The sample period is February 1987 through May 2009 for the three series.

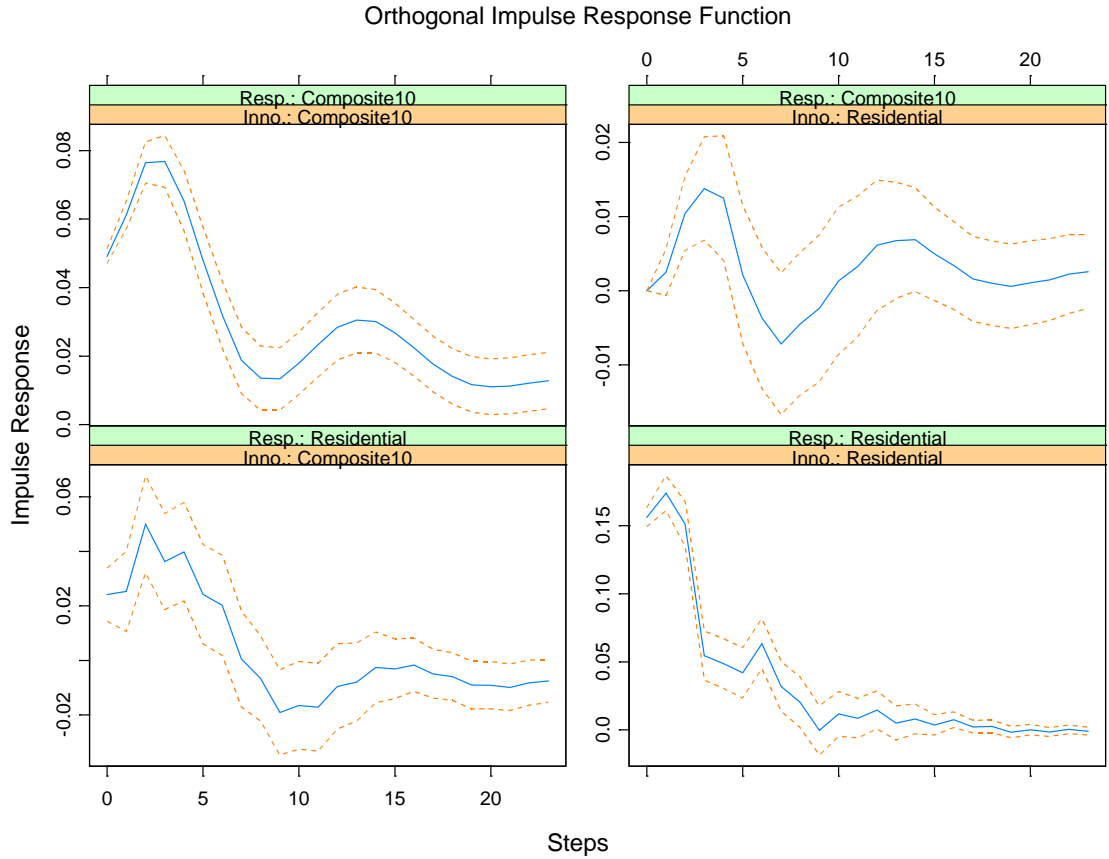

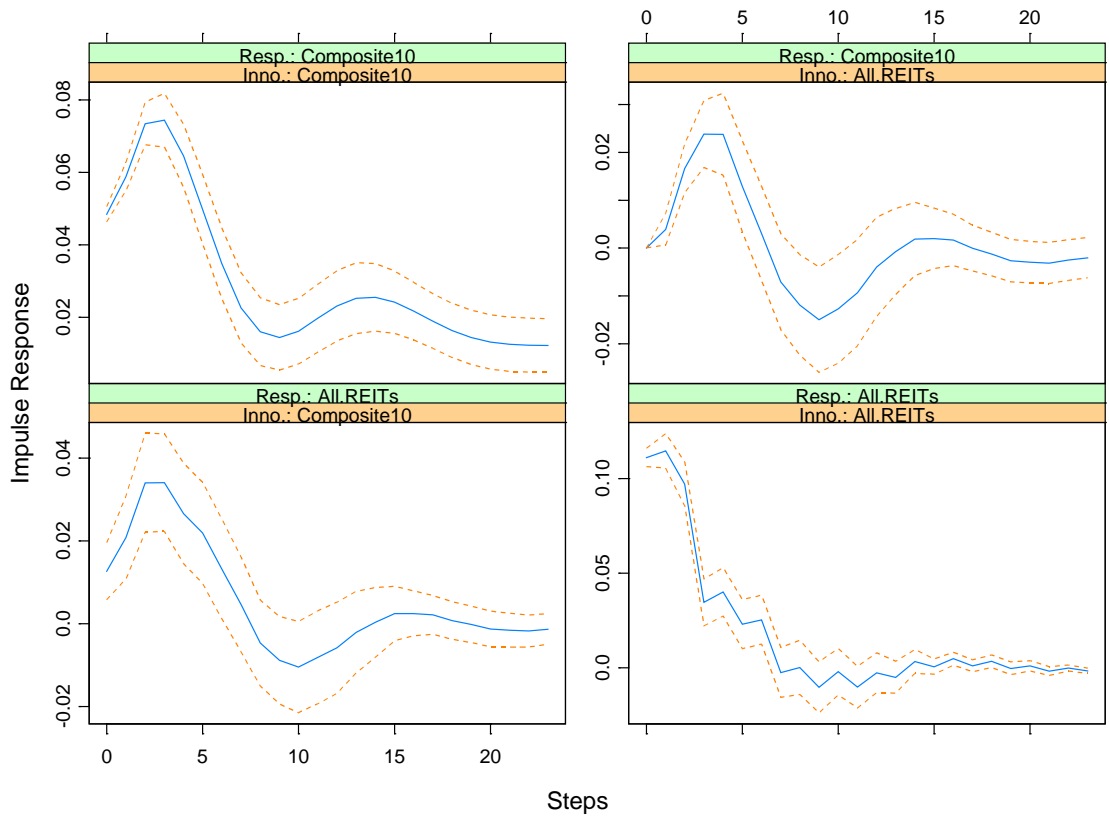

Figure 9. Forecast Error Variance Decompositions for Real Estate Returns

Forecast error variance decompositions and asymptotic standard errors from the VAR model for a sample of series are presented: All REITs Value Weighted Index, REITs Residential Property Value Weighted Index respectively with the SCS Composite-10 index. The sample period is February 1987 through May 2009 for the three series.

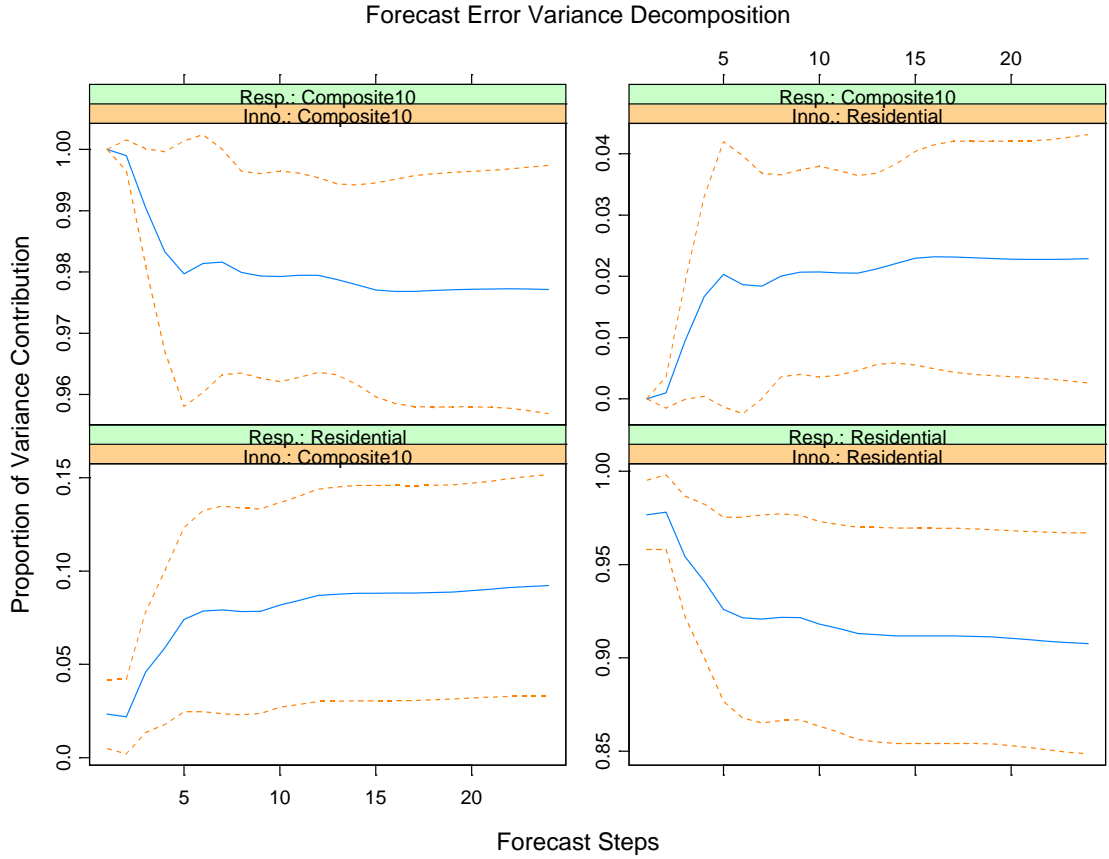

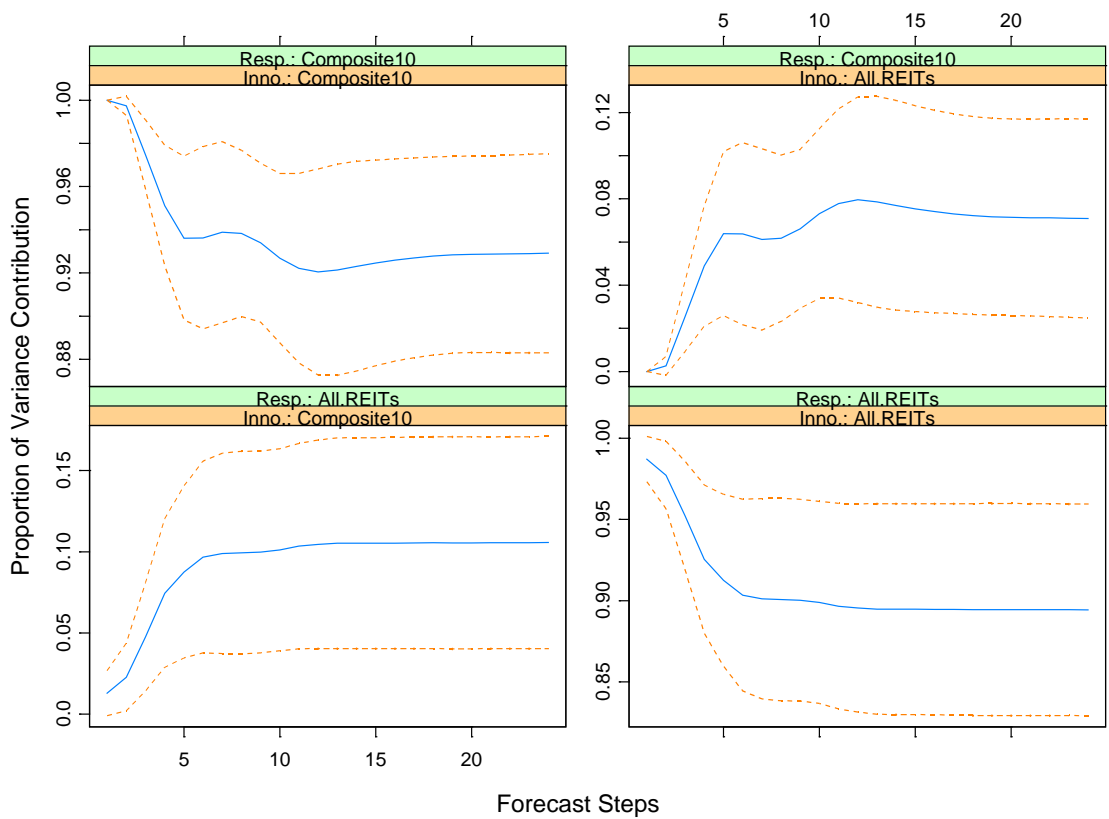